\documentclass[aps,pre,reprint,groupedaddress]{revtex4-2}

\usepackage{graphicx,amsmath}
\usepackage{hyperref}
\hypersetup{colorlinks=true, citecolor=blue, urlcolor=blue, linkcolor=blue}
\usepackage{xcolor}

\begin{document}


\title{Shear viscosity and Stokes-Einstein violation in supercooled light and heavy water}


\author{Pierre Ragueneau}
\author{Fr\'ed\'eric Caupin}
\email[]{frederic.caupin@univ-lyon1.fr}
\author{Bruno Issenmann}
\email[]{bruno.issenmann@univ-lyon1.fr}
\affiliation{Institut Lumi\`ere Mati\`ere, Universit\'e de Lyon, Universit\'e Claude Bernard Lyon 1, CNRS, F-69622, Villeurbanne, France}

\date{\today}

\begin{abstract}
We report shear viscosity of heavy water supercooled $33\,\mathrm{K}$ below its melting point, revealing a 15-fold increase compared to room temperature. We also confirm our previous data for the viscosity of supercooled light water, and reach a better accuracy. Our measurements, based on the spontaneous Brownian motion of $350\,\mathrm{nm}$ spheres, disagree at the lowest temperature with the only other available data, based on Poiseuille flow in a narrow capillary, which may have been biased by electro-osmotic effects. The viscosity ratio between the two isotopes reaches 2.2 at the lowest temperature. A companion Letter [F.~Caupin, P.~Ragueneau, and B.~Issenmann, arXiv:2112.09010] discusses this giant dynamic isotopic effect. Here we provide a detailed description of the experiment and its analysis. We review the literature data about dynamic properties of water (viscosity, self-diffusion coefficient, and rotational correlation time), discuss their temperature dependence and compare their decoupling in the two isotopes.
\end{abstract}


\maketitle



\section{Introduction}

When a liquid can be measured during cooling to its glass transition temperature $T_{g}$, its shear viscosity is observed to increase tremendously, sometimes over more than 14 decades. One of the definitions of $T_{g}$ is the temperature at which viscosity reaches $10^{12}\,\mathrm{Pa\,s}$. The detailed temperature dependence of viscosity varies between liquids. Some, such as silica, exhibit a nearly Arrhenius behavior, whereas others, such as ortho-terphenyl, have a viscosity varying much faster than an Arrhenius law. The former are called strong liquids, and the latter fragile~\cite{angell_formation_1995}. The glass transition is also characterized by heat capacity measurements. It is found that the relative temperature width of this calorimetric glass transition is large in strong liquids, and small in fragile ones~\cite{ito_thermodynamic_1999}.

Water shows a hybrid behavior. Although its viscosity can be measured over a relatively modest range due to crystallization above $230\,\mathrm{K}$~\cite{goy_shrinking_2018}, available data shows a super-Arrhenius temperature dependence, typical of a fragile liquid. In contrast, the relative temperature width of its calorimetric glass transition (measured when heating low density amorphous ice) ranks water among the strongest liquids. This has led to the suggestion that water undergoes a fragile-to-strong transition~\cite{ito_thermodynamic_1999}, unfortunately lying in a temperature range not accessible to experiments.

Another puzzle with water as a glassformer is the early decoupling of its dynamic properties. Viscosity $\eta$, self-diffusion coefficient $D_{s}$, and rotational correlation time $\tau_{\theta}$ (see Supplemental Material~\cite{SM} for a definition of $\tau_{\theta}$ and its  discussion in the case of water) are usually tightly linked in liquids at high temperature, through the Stokes-Einstein (SE) and Stokes-Einstein-Debye (SED) relations, which respectively imply that $D_{s}\eta/T$ and $\eta/(T\tau_{\theta})$ do not depend on temperature $T$. For usual fragile glassformers, $D_{s}\eta/T$ starts increasing when $T$ decreases below around $1.3\,T_{g}$~\cite{chang_heterogeneity_1997}. In contrast, measurements on ortho-terphenyl have shown that $\eta/(T\tau_{\theta})$ remains fairly constant over 14 decades of viscosity~\cite{cicerone_photobleaching_1993}. Water exhibits a qualitatively similar behavior, with a violation of the SE relation stronger than that of the SED relation~\cite{dehaoui_viscosity_2015}. However, the SE violation already starts at room temperature, more than $2\,T_{g}$.

To investigate further the SE and SED violations in water, accurate dynamic data in the supercooled region is needed. While available for $D_{s}$~\cite{price_self-diffusion_1999} and $\tau_{\theta}$~\cite{qvist_rotational_2012}, data for $\eta$ is scarce. Two studies reported viscosity of deeply supercooled water~\cite{hallett_temperature_1963,osipov_shear_1977}, but they disagree at the lowest temperatures. They were both based on Poiseuille flow, but used capillaries with very different diameters. This led us to perform measurements with an independent technique, using Brownian motion of spherical probes to obtain $\eta$ without steady flow~\cite{dehaoui_viscosity_2015}. We could thus measure a 14-fold increase of viscosity from $293.15$ to $239.15\,\mathrm{K}$, and revealed a bias in the previous Poiseuille-flow study with the smallest capillary~\cite{osipov_shear_1977}, which we attributed to electro-osmotic effects. Our data thus confirmed the fragile behaviour of water's viscosity, the early violation of the SE relation, and a milder violation of the SED relation~\cite{dehaoui_viscosity_2015}.

Because the hydrogen bond, at the heart of water's anomalies, is strongly affected by isotopic substitution of hydrogen with deuterium~\cite{ceriotti_nuclear_2016}, comparison between light and heavy water can provide new insight. Trading H$_2$O for D$_2$O represents a larger relative change in the moment of inertia of a molecule than in its mass. Therefore, one may expect that translational and rotational properties are affected in different ways. Unfortunately, whereas $D_{s}$~\cite{price_self-diffusion_1999} and $\tau_{\theta}$~\cite{qvist_rotational_2012} data for supercooled D$_2$O are available, the only viscosity data is from Ref.~\cite{osipov_shear_1977}, which we proved to be biased in the case of H$_2$O. It appears therefore necessary to produce reliable data for the viscosity of supercooled D$_2$O. To this end, we have applied the Brownian motion method to heavy water. We report here values down to $243.7\,\mathrm{K}$, $33.3\,\mathrm{K}$ below the melting point.

During this study, we were able to improve our measurement procedure, so that we also report new data for light water which agree with our previous results, but have better accuracy. Viscosity of supercooled water is a key parameter in the spreading of cold droplets on surfaces~\cite{maitra_supercooled_2014,schremb_normal_2018,ding_how_2018}, making its precise knowledge relevant to the phenomenon of icing of roads and aircrafts.

The paper is organized as follows. Experimental details are presented in Section~\ref{Ragueneau_ExperimentalDetails} and results in Section~\ref{Ragueneau_Results}. Section~\ref{RagueneauLiterature} reviews the literature data on dynamic quantities of water (viscosity, self-diffusion coefficient and rotational correlation time). Their temperature dependence is analyzed in Section~\ref{Ragueneau_Discussion}, which also discusses the SE and SED violation. Concluding remarks are given in Section~\ref{sec:conclusion}.

\section{Materials and Methods}

\label{Ragueneau_ExperimentalDetails}

\subsection{Differential Dynamic Microscopy}

\begin{figure}
	\includegraphics[width=8.6cm]{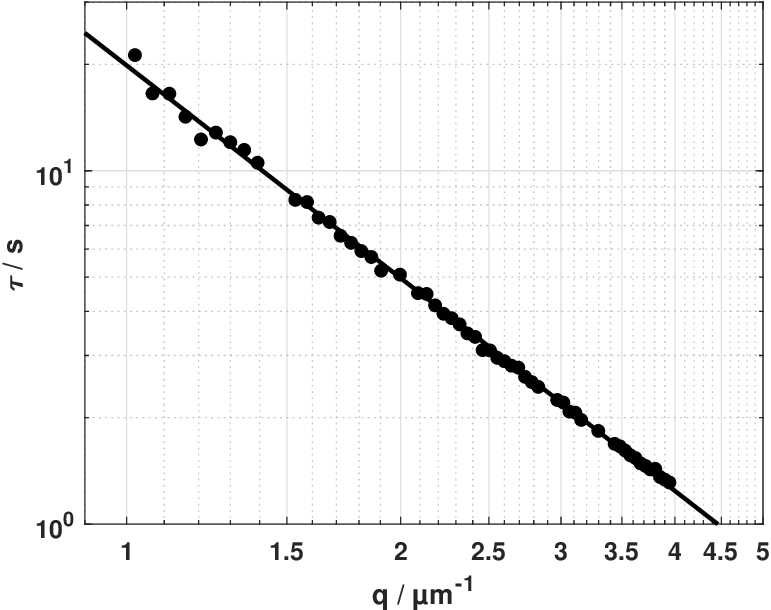}
	\caption{\label{Ragueneau_tauq}Decorrelation time as a function of wave vector $q$ in heavy water (mass fraction D$_2$O $97\%$) at $-31^{\circ}$C. The line is a fit with $\tau = 1/(Dq^2)$, which yields $D=5.03\,10^{-2}\,\rm{{\mu m}^2\,s^{-1}}$.}
\end{figure}

To measure the viscosity of supercooled water, we use Differential Dynamic Microscopy (DDM) ~\cite{cerbino_differential_2008,giavazzi_scattering_2009} as described in Ref.~\cite{dehaoui_viscosity_2015}. In brief, monodisperse Brownian spheres (Duke Scientific, diameter $2r=350$~nm) are dispersed in the liquid of interest. The samples are observed with a microscope (Zeiss Axioscope), equipped with a long working distance x100 objective (Mitutoyo M-Plan APO) and a CCD camera (Allied Vision Prosilica $1024\times 1024$) with a frame rate up to 112~fps. The Brownian spheres are smaller than the resolution limit of the microscope, but scatter light when illuminated in transmitted light mode. As recommended in Ref.~\cite{cerbino_differential_2008}, the numerical aperture of the condenser is reduced to its minimum to decrease the incoherence parameter. 
Due to the temporal evolution of the light scattered by the colloids, the difference between two images taken at times $t_1$ and $t_2$ shows a granularity which increases with $|t_2 - t_1|$. Analysis of the spatial Fourier transform of image differences yields a decorrelation time $\tau$ as a function of wave vector modulus $q$. As explained in Refs. ~\cite{cerbino_differential_2008,giavazzi_scattering_2009} and shown in Fig.~\ref{Ragueneau_tauq}, for Brownian motion, $\tau = 1/(D q^2)$, where $D$ is the diffusion coefficient of the colloids.

The shear viscosity $\eta$ follows from the Stokes-Einstein equation for a Brownian sphere:
\begin{equation}
	\eta=\frac{k_B T}{6\pi rD}\,,
	\label{StokesEinstein}
\end{equation}
where $k_B$ is the Boltzmann constant and $T$ the temperature.

This technique is particularly suitable for measuring viscosity at deeply supercooled conditions, because it involves small volumes and does not induce strong perturbations in the fluid, such as steady flow or shear.

\subsection{The colloidal suspension}

\subsubsection{H$_2$O suspension}

The commercial colloidal suspension contains monodisperse polystyrene spheres suspended in water. To avoid the presence of possible contaminants like surfactants that could be included in the suspension, the suspension is first rinsed using the following procedure. First, the commercial suspension is diluted with ultrapure water (Direct-Q3, UV, Millipore) to obtain a $0.01\%$ mass fraction of particles. Then $2$~mL of the diluted suspension are centrifugated. The colloids fall at the bottom of the suspension and the supernatant liquid, including a large part of the possible contaminants, is removed with a pipette. Finally, ultrapure water is added to obtain a suspension with $0.01\%$ mass fraction of colloids. The whole procedure is carried out twice.

\subsubsection{D$_2$O suspension}

The samples for the measurements of the viscosity of D$_2$O are prepared in a different way since the colloids are less dense than heavy water and cannot be centrifugated. To remove possible contaminants, the commercial suspension is diluted with ultrapure H$_2$O (to reach $0.1\%$ mass fraction of colloids). This suspension is rinsed twice using the same procedure as above. Then it is centrifugated a third time and as much H$_2$O as possible is removed and replaced by D$_2$O ($\ge 99.9\%$, Eurisotop), to obtain a suspension whose mass fraction in colloids is $0.1\%$. Finally, this suspension is diluted ten times in D$_2$O to obtain the desired $0.01\%$ mass fraction of particles. 

\subsection{Experimental setup}

\subsubsection{Sample preparation\label{sec:sample}}

\begin{figure}
	\includegraphics[width=8.6cm]{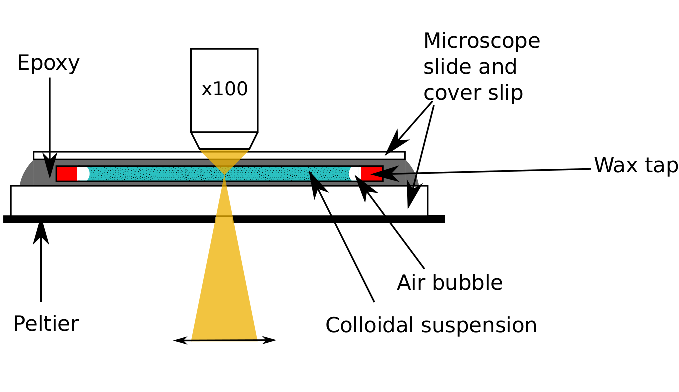}
	\caption{\label{Ragueneau_GluedCapillary}Side view of the sample placed in the temperature-controlled stage of the microscope.} 
\end{figure}

After rinsing, the suspension is placed in a borosilicate glass capillary with rectangular cross-section (Vitrotubes, internal dimensions $10\times0.2\times0.02$~mm, wall thickness around $0.01$~mm). The capillary is then sealed by dipping its ends into fused wax (Hampton Research). Before sealing, we wait for evaporation of a quarter of the suspension, thus leaving air bubbles in order to avoid a pressure increase due to the expansion of water upon cooling. The sealed capillary is placed on a standard microscope slide (thickness $1$~mm) together with two capillaries containing respectively pure water and pure dodecene for temperature calibration (see \S~\ref{TemperatureCalibrationPrinciple}). The three capillaries are covered with epoxy. Finally, a microscope cover slip is put on top of them and the sample is left drying for 15 minutes under a 230~g weight. Fig.~\ref{Ragueneau_GluedCapillary} shows a sketch of the final setup. The temperature calibration capillaries are omitted in this figure for clarity. 
The sample is placed in a thermal stage (Linkam LTS120 Peltier system) to control its temperature down to $233\mathrm{K}$; silica gel is added in the stage chamber to avoid moisture condensation on the cold microscope cover slip. The thermal stage is fixed on the microscope for DDM measurement.  

\subsubsection{Acquisition parameters}

DDM requires a fast enough acquisition to capture particle displacements of the order of their size between successive frames. As the mean square displacement of a Brownian particle during time $t$ scales as $\sqrt{D t}$, the acquisition frequency should scale as $D$, which is inversely proportional to the shear viscosity $\eta$ (Eq.~\ref{StokesEinstein}). Therefore, as a run is performed for a series of decreasing temperatures, corresponding to increasing $\eta$ values, the frame rate $\mbox{fps}$ of the camera is successively decreased as $\mbox{fps} \propto 1/\eta$, using an approximate extrapolated value of $\eta$ for this calculation. Typical data consist of a sequence of 1500 square images (1024x1024 pixels), acquired at $10-112\,\mathrm{fps}$ depending on temperature, with $8\,\mathrm{ms}$ exposure time.

\subsection{Temperature calibration}

As the viscosity of water is a steep function of temperature in the supercooled region~\cite{dehaoui_viscosity_2015}, it is critical to have an accurate value for the sample temperature, which might differ from the nominal temperature of the thermal stage. For example, complete melting of ultrapure ice is observed at a nominal temperature of $-0.3\pm0.1^{\circ}$C, which reveals the existence of temperature gradients across the sample. We found a negligible change in the melting point of ice when replacing the bottom microscope slide by a thinner one. Therefore, the temperature gradient is rather a lateral one, due to the hole in the center of the Peltier element, which allows illuminating the sample. We note that the quality of the thermal contact between the polished heating/cooling element of the stage, the microscope slide, and the capillaries may affect this temperature gradient. We checked this by measuring the melting point of water and undecane in around ten capillaries prepared as described in Section~\ref{sec:sample}. We found a dispersion of the melting temperatures lower than the reading resolution ($0.1^{\circ}$C). This shows that immersing the capillaries in epoxy allows achieving a reproducible thermal contact and reducing temperature variability. In the remainder of this section, we describe how we calibrated the temperature in the observed section of the capillary using the melting point of pure chemicals.

\subsubsection{Melting point determination of calibration chemicals}
\label{CalibrationMeltingPoint}
We selected a series of pure chemicals covering the temperature range of our experiments and measured their actual melting temperature $T_{m,\mathrm{act}}$ as follows. $16\,\mathrm{\mu L}$ of the chemical are put in a Pasteur pipette, previously fused-welded at one end. A platinum resistor (Pt-$100~\Omega$, $1/3$ DIN, previously calibrated in melting ice) is immersed in the liquid and its resistance measured with a digital multimeter (Tektronix DMM4050) in 4-wires mode. The Pasteur pipette is immersed in a thermal bath (Julabo FP89-HL) whose temperature is lowered until complete crystallization of the chemical. Then the bath temperature is ramped up to room temperature. Figure~\ref{ResultatCalibrationTemperature} (inset) displays a typical temperature trace of the Pt-$100$ resistor, from which $T_{m,\mathrm{act}}$ is obtained. The results for 4 chemicals are given in Table~\ref{tab_MeltingPoints}.

\begin{table*}
\caption{\label{tab_MeltingPoints} Melting points of the 4 chemicals used for temperature calibration (see text for details). Uncertainties correspond to a $68\%$ confidence interval.}
\begin{ruledtabular}
\begin{tabular}{lccc}
Chemical 				& Actual melting point $T_{m,\mathrm{act}}$ ($\rm{^o}$C)	&	Tabulated melting point~\cite{noauthor_section_2014} ($\rm{^o}$C) 	&	$T_{m,\mathrm{Linkam}}$ ($\rm{^o}$C)\\
Dodecene				&								$-35.18\pm0.02$													&										-35.19																	&						$-36.6\pm0.1$						\\
Undecane				&								$-25.92\pm0.06$													&										-25.54																	&						$-26.8\pm0.1$						\\
Ultrapure water	&								 0																			&											0																			&						$-0.4\pm0.1$						\\
Heptadecane			&								$21.1\pm0.3$														&										21.97																		&						$21.2\pm0.1$						\\
\end{tabular}
\end{ruledtabular}
\end{table*}

\subsubsection{Principle of the calibration with pure chemicals}

\label{TemperatureCalibrationPrinciple}

To calibrate the temperature of the sample in the observed section of the capillary, we make the same samples as described above, only replacing the colloidal suspension inside the capillary by a pure chemical. After being placed in the thermal stage, the sample is cooled until it crystallizes. Then, by slow step-wise heating, we determine the nominal stage temperature $T_{m,\mathrm{Linkam}}$ at which a crystal located at the center of the observation area grows or remains stable at $T_{m,\mathrm{Linkam}}-0.1$~K, while it completely melts at $T_{m,\mathrm{Linkam}}$. The precision on $T_{m,\mathrm{Linkam}}$ is $\pm0.1^{\circ}$C. Comparison of this measurement with the actual melting temperature of the chemical provides a calibration of the thermal stage.

\begin{figure}
	\includegraphics[width=8.6cm]{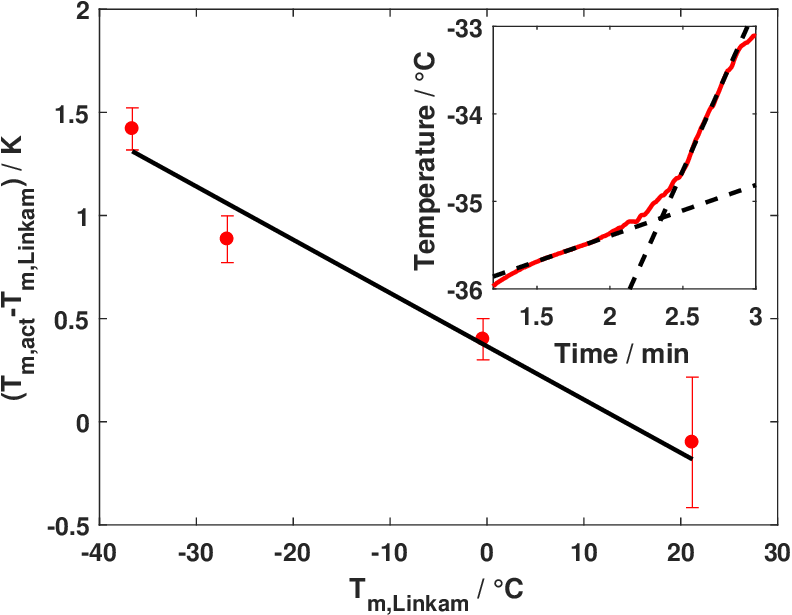}
	\caption{\label{ResultatCalibrationTemperature}$\Delta T=T_{m,\mathrm{act}}-T_{m,\mathrm{Linkam}}$ as a function of $T_{m,\mathrm{Linkam}}$ for the 4 chemicals listed in Table~\ref{tab_MeltingPoints}. The line is a linear least-$\chi_2$ fit. Inset: Determination of the melting point of dodecene during heating in the thermal bath: $ T_{m,\mathrm{act}}$ is given by the intersection of the two straight lines.}
\end{figure}

Figure~\ref{ResultatCalibrationTemperature} shows the difference $\Delta T=T_{m,\mathrm{act}}-T_{m,\mathrm{Linkam}}$ as a function of $T_{m,\mathrm{Linkam}}$. In the temperature range of our experiment, $\Delta T$ is a linear function of $T_{m,\mathrm{Linkam}}$. However, the exact function may vary with sample, for instance due to a change in room temperature or a different contact between the Peltier element and the microscope slide. To avoid this issue, we performed an \emph{in situ} temperature calibration for each sample, thanks to the two capillaries added next to the capillary containing the colloidal suspension (see Section~\ref{sec:sample}). One is filled with dodecene, and the other with ultrapure water. The measurement of the corresponding melting points just after the DDM run allows a quick calibration of the linear relationship between $\Delta T$ and $T_{m,\mathrm{Linkam}}$ for that run, thus giving accurate values for the sample temperatures during the run.

\subsection{Isotopic fraction}

An additional difficulty of the experiments with D$_2$O is isotopic purity. The sample can be contaminated by light water during preparation, due to isotopic exchange with ambient water vapor and to the light water present in the commercial colloidal solution. To minimize contamination, the heavy water bottle and prepared colloidal solution are stored in a nitrogen container. All sample preparations are performed under a steady nitrogen flux. Despite all these precautions, the D$_2$O fraction in the resulting sample was less than in the initial solution. The measured viscosity $\eta$ is then the one of an isotopic mixture, whose D$_2$O mole fraction $x$ needs to be determined.

We obtained $x$ from the melting point of the mixture. After crystallization upon cooling, we warm up the sample in the stage until complete melting is observed under the microscope. The melting point $T_\mathrm{m}$ (in $^{\circ}$C) of D$_2$O-H$_2$O mixtures is $T_\mathrm{m}=4.213x-0.411x^2$~\cite{mer_freezing_1934}. The accuracy on $T_m$ is around $0.15\,\mathrm{K}$, resulting in an absolute accuracy of $0.04$ on $x$. This measurement is performed at the end of a run, together with the temperature calibration, since crystallization of the solution leads to aggregation of the colloids. Typical values of $x$ ranged from $0.47$ to $0.97$.

\subsection{Viscosity calculation for H$_2$O\label{sec:etacalcH2O}}

For each run, a reference diffusion coefficient is measured at $T_0=293.15$~K, where the viscosity of pure light water is known with the best accuracy and precision ~\cite{huber_new_2009}: $\eta_0(T_0)=1.0016\pm 0.0017$~mPa.s. Here $T_0$ is the nominal temperature of the stage. At this temperature, the difference between $T_0$ and the actual one is small, and the viscosity has only a small temperature dependence, so that a correction of $T_0$ is not necessary.

DDM gives $D(T)$, the diffusion coefficient ot the Brownian spheres as a function of temperature. To convert these values into viscosity $\eta_0(T)$ of water, we take the ratio of the Stokes-Einstein equations (Eq.~\ref{StokesEinstein}) at temperature $T$ and at the reference temperature $T_0$, which yields:
\begin{equation}
	\eta_0(T)=\eta_0(T_0)\frac{T}{T_0}\frac{D(T_0)}{D(T)} \, .
	\label{StokesEinsteinRenormalise}
\end{equation}

As explained before~\cite{dehaoui_viscosity_2015}, this relative measurement avoids possible effects of the capillary walls and of the electric double layer around the colloids on their diffusion coefficient. Knowledge of the sphere radius $r$ is not needed, as it cancels out when taking the ratio of the Stokes-Einstein equations, because the expansion coefficient of polystyrene is negligible in this temperature range.

Viscosity values for H$_2$O are given for each run in the Supplementary Information. A run corresponds to one sample, containing a capillary with colloids suspended in water, and two capillaries containing respectively dodecene and pure water to calibrate \emph{in situ} the temperature as described above. For most temperatures, the viscosity was measured several times (often 3). 

In our previous work~\cite{dehaoui_viscosity_2015}, $D(T_0)$ was measured in 12 independent samples which resulted in a $2.5\%$ standard deviation. However, the standard deviation of $D(T_0)$ measured 10 times on the same sample is around $1.5\%$ only. In the present work, to improve the precision of our data, we applied Eq.~\ref{StokesEinsteinRenormalise} to each sample by measuring $D(T_0)$ 10 times on this sample before performing experiments at other temperatures, instead of rescaling the data for different samples by the same reference measurement averaged on several samples. Therefore, we now take for the intrinsic relative uncertainty (1 SD) on $\eta$ $1.5\%$. The temperature uncertainty $\delta T$ also contributes to the uncertainty on viscosity. To take it into account, we first least-square fit the viscosity values by the Speedy-Angell law
\begin{equation}
\eta(T) = \eta_0 \left(\frac{T}{T_{s}}-1 \right)^{-\gamma} \, ,
\label{eq:etaSpeedyAngell}
\end{equation}
and calculate the total relative uncertainty (1 SD) at temperature $T$ as $\sqrt{(0.015^2+(\delta T\times \gamma/(T-T_s))^2}$. The resulting uncertainty ranges from $1.5\%$ at the highest temperature to $2.2\%$ at the lowest temperature.

\subsection{Viscosity calculation for D$_2$O}

Viscosity values for D$_2$O were calculated in the same way as for H$_2$O based on a reference viscosity $\eta(x,T_0)$ for each sample with D$_2$O mole fraction $x$, where $T_0=293.15$~K the same reference temperature as above. At $T_0$, viscosity is well described by a linear relation ~\cite{kestin_viscosity_1985} :
\begin{equation}
	\eta(x,T_0)=(1-x)\eta(0,T_0) + x\eta(1,T_0) \, .
	\label{Ragueneau_ExtrapolationD2O}
\end{equation}

The viscosity $\eta(1,T_0)$ of heavy water at the reference temperature $T_0$ was measured by Millero~\cite{millero_density_1971}. Millero calibrated his measurements using the reference value for the viscosity of H$_2$O at $T_0$ known at that time, that was $1.0020\,\mathrm{mPa\,s}$. This reference value was since re-evaluated to $\eta(0,T_0) = 1.0016 \pm 0.0017\,\mathrm{mPa\,s}$~\cite{huber_new_2009}. Therefore, we recalculated Millero's values and propagated the errors of Millero and $\eta(0,T_0)$ to obtain $\eta(1,T_0)=1.2466 \pm 0.0021\,\mathrm{mPa\,s}$.

The viscosity $\eta(x,T)$ of the sample are then calibrated using the reference value $\eta(x,T_0)$ in Eq.~\ref{StokesEinsteinRenormalise}. We thus obtain viscosity values for the isotopic mixtures at each temperatures. To extrapolate the data to pure D$_2$O, we still need to check that the linear Eq.~\ref{Ragueneau_ExtrapolationD2O} remains valid far in the supercooled region. To do so, all the data were binned into $0.1~K$ intervals and fitted by 
\begin{equation}
	\eta ( x,T ) = x \,\eta(1,T) + a\,( 1 - x ) \, .
	\label{Ragueneau_FitLineaireEta}
\end{equation}
Smoothed viscosity values for pure H$_2$O calculated from Eq.~\ref{eq:etaSpeedyAngell} with the best-fit parameters listed in table~\ref{RagueneauFitsPower} are included in the fitted data at all temperatures. The uncertainty on $x$ is projected on the vertical axis and combined to the uncertainty on $\eta$.

\begin{figure}
  \includegraphics[width=8.6cm]{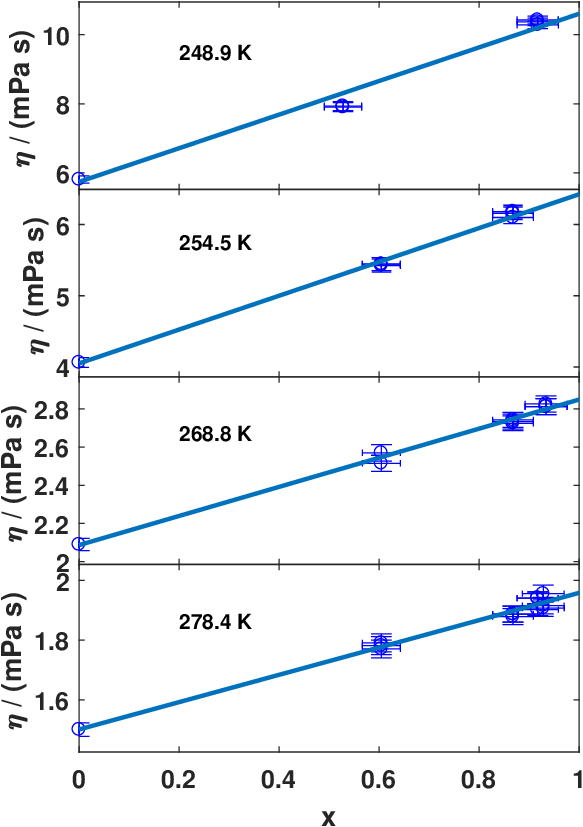}  
  \caption{Shear viscosity $\eta$ as a function of D$_2$O molar fraction $x$ for various temperatures ($\pm 0.05$~K). The lines are least-$\chi_2$ fit with Eq.~\ref{Ragueneau_FitLineaireEta}.}
  \label{Ragueneau_EtaX}
\end{figure}

The fitting parameters are $a$, which agrees well with the measured values of the viscosity of pure H$_2$O, and $\eta(1,T)$, which is thus taken as the viscosity of pure D$_2$O. Typical examples are displayed in Fig.~\ref{Ragueneau_EtaX}, showing that the linear relation remains valid over the whole temperature range of our study.

\begin{figure}
	\includegraphics[width=8.6cm]{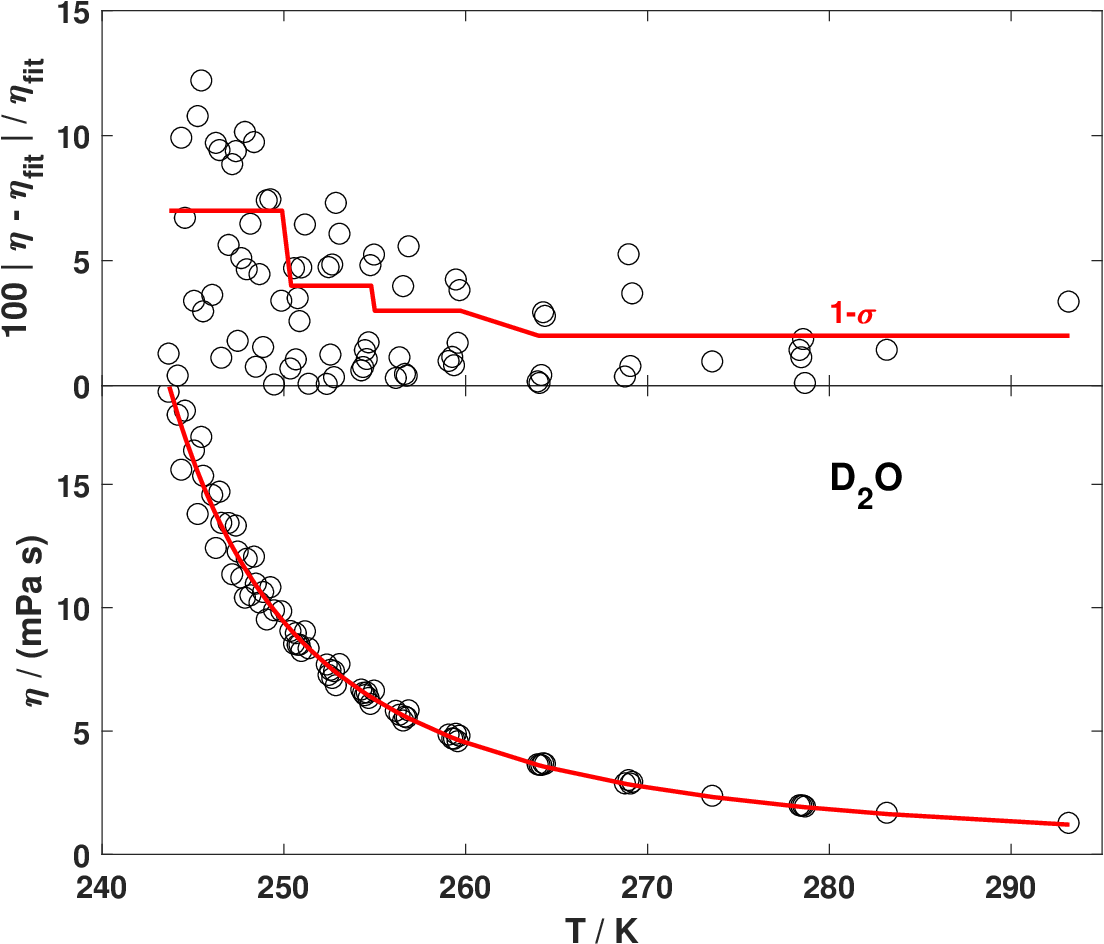}
	\caption{\label{Ragueneau_Dispersion_D2O} Bottom panel: raw viscosity data for pure D$_2$O. The red curve is a least-square fit with the Speedy-Angell power-law (Eq.~\ref{eq:etaSpeedyAngell}). Top panel: Relative deviation between data and fit. The $1-\sigma$ relative uncertainty corresponds to a $68\%$ confidence interval.}
\end{figure}

As the sources of uncertainty in this measurement is more complex than in light water (in particular due to the uncertainty on the mass fraction $x$ of heavy water), we could not use the same procedure to estimate the final uncertainty. We first least-square fit the raw data by a Speedy-Angell law (Fig.~\ref{Ragueneau_Dispersion_D2O}, bottom panel). The deviation of the experimental data from the fitting equation is shown in the top panel. It can be seen that the data scatter increases at low temperatures. To be conservative, 4 different temperature ranges were considered, and the relative uncertainty was calculated as the standard deviation of the data in each temperature range. The values are given in Table~\ref{Ragueneau_BarresDErreurD2O}.

\begin{table}
	\caption{\label{Ragueneau_BarresDErreurD2O}$1-\sigma$ relative uncertainties on the measurements of $\eta_{\rm{D_2O}}$ as a function of temperature.}
		\begin{ruledtabular}
			\begin{tabular}{cc}
				Temperature range 				& $\frac{d\eta_{\rm{D_2O}}(T)}{\eta_{\rm{D_2O}}(T)}$ ($\%$)		\\
				$ T \ge 260 $~K 						& 		2				\\
				$ 255$~K $\le T < 260 $~K 	& 		3				\\
				$ 250$~K $\le T < 255 $~K	  & 		4				\\
				$  T < 250$~K 							& 		7				\\
			\end{tabular}
		\end{ruledtabular}
\end{table}

\section{Results}

\label{Ragueneau_Results}

\subsection{Improved viscosity values for H$_2$O}

\label{Ragueneau_Results_H2O}

Figure~\ref{Ragueneau_FigH2O} displays the raw values of the viscosity of light water (listed in Supplementary Information) and their least-$\chi_2$ fit by a Speedy-Angell law with $\eta_0 = 0.13746\,\mathrm{mPa\,s}$, $T_{s} = 225.9151\,\mathrm{K}$ and $\gamma = 1.6383$. The resulting smoothed values $\eta_{\rm{smoothed,H_2O}}(T)$ are given in Table~\ref{Ragueneau_Smoothed}. The fit reduced residuals for individual measurements are also shown in the top panel of Fig.~\ref{Ragueneau_FigH2O}: they are consistent with our estimate of the measurement uncertainty (Section~\ref{sec:etacalcH2O}).

\begin{figure}
	\includegraphics[width=8.6cm]{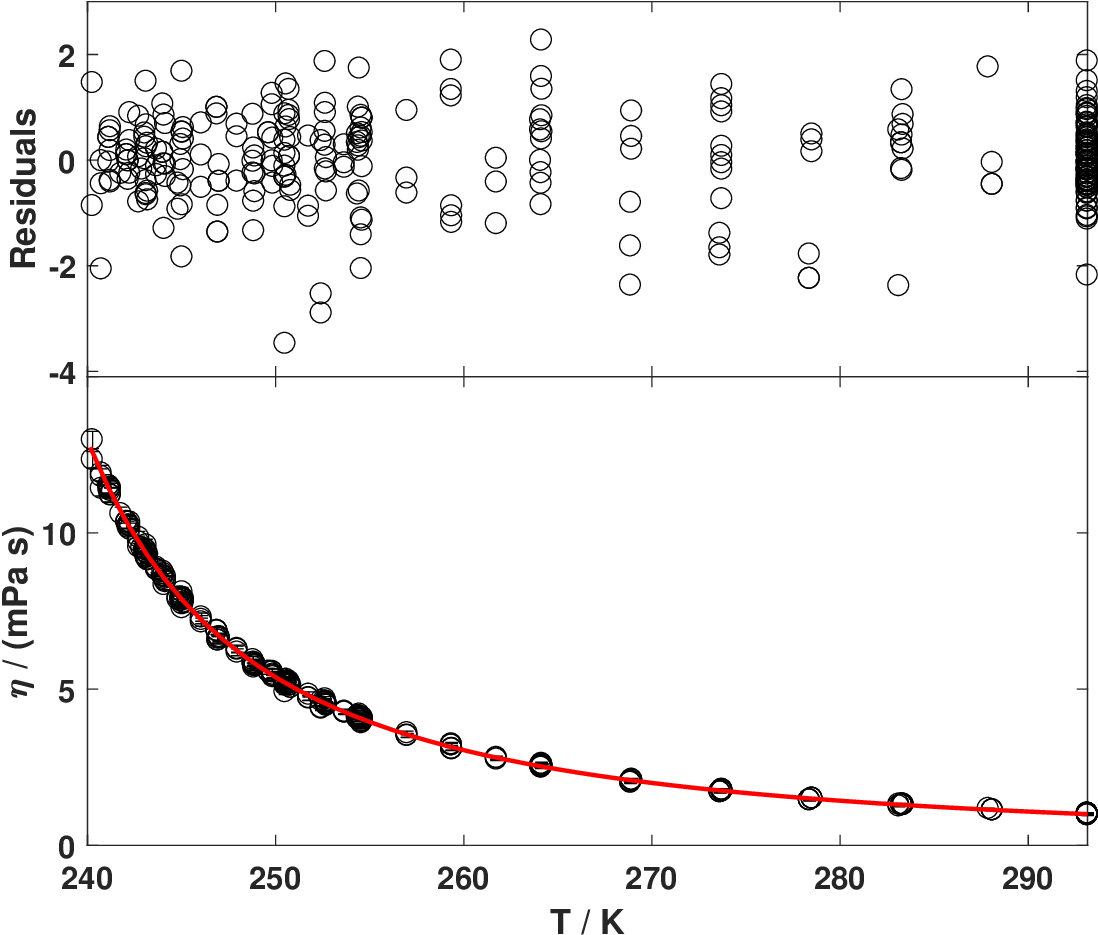}
	\caption{\label{Ragueneau_FigH2O} Lower panel: viscosity of H$_2$O as a function of temperature. The red curve is $\eta_{\rm{smoothed,H_2O}}(T)$. Upper panel: reduced residuals $[\eta(T)-\eta{\rm{smoothed}}(T)]/\sigma(T)$.}
\end{figure}

\begin{figure}
	\includegraphics[width=8.6cm]{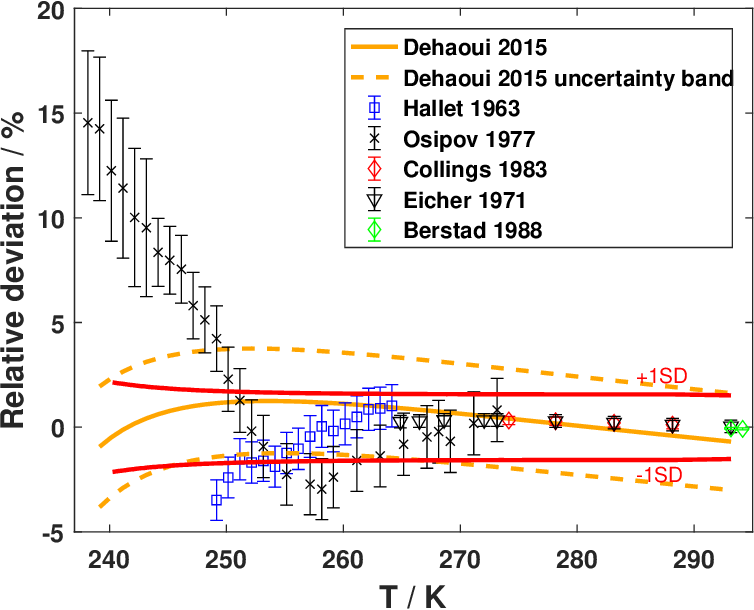}
	\caption{\label{Ragueneau_CompaLittH2O} Relative deviation of literature data~\cite{dehaoui_viscosity_2015,hallett_temperature_1963,eicher_high-precision_1971,osipov_shear_1977,collings_high_1983,berstad_accurate_1988} from our results $\eta_{\rm{smoothed,H_20}}$ (Table~\ref{Ragueneau_Smoothed}). The curves show the uncertainty band ($\pm 1$ standard deviation) for our previous data~\cite{dehaoui_viscosity_2015} (dashed orange) and the present ones (solid red).}
\end{figure}

Figure~\ref{Ragueneau_CompaLittH2O} compares the present data to previously published data sets~\cite{dehaoui_viscosity_2015,hallett_temperature_1963,eicher_high-precision_1971,osipov_shear_1977,collings_high_1983,berstad_accurate_1988}. We confirm with a better precision our previous results~\cite{dehaoui_viscosity_2015}, and, as already explained therein, we confirm the agreement with Collings~\cite{collings_high_1983}, Berstad~\cite{berstad_accurate_1988}, Hallet~\cite{hallett_temperature_1963} and Eicher~\cite{eicher_high-precision_1971} (as corrected by Kestin~\cite{kestin_viscosity_1978}), except for a slight deviation from Hallet at the lowest temperatures. We confirm the large deviation with Osipov's data~\cite{osipov_shear_1977} that we previously attributed to a possible electro-osmosis effect that could have biased their results~\cite{dehaoui_viscosity_2015}.

\subsection{Viscosity values for D$_2$O}

\label{Ragueneau_Results_D2O}

Figure~\ref{Ragueneau_Dispersion_D2O} displays raw viscosity values for D$_2$O (listed in the Supplementary Information). Using uncertainties from Table~\ref{Ragueneau_BarresDErreurD2O}, they are least-$\chi_2$ fitted by a Speedy-Angell law, which yields the following best-fit parameters: $\eta_0=1.3212.10^{-4}$~Pa.s, $T_s=230.9681$~K and $\gamma=1.7061$. The smoothed values $\eta_{\rm{smoothed,D_2O}}(T)$ calculated using this fit are given in Table~\ref{Ragueneau_Smoothed}.

\begin{figure}
	\includegraphics[width=8.6cm]{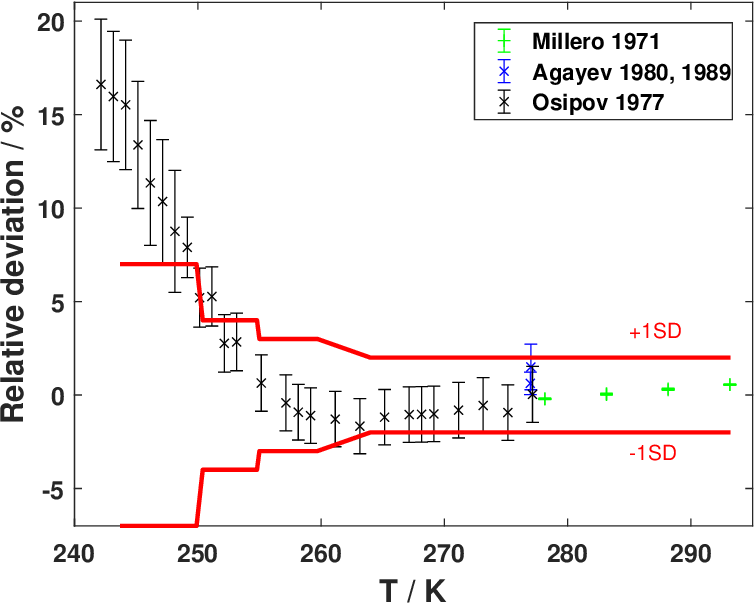}
	\caption{\label{Ragueneau_CompaLittD2O} Relative deviation of literature data~\cite{millero_density_1971,osipov_shear_1977,agayev_experimental_1980,agayev_heavy-water_1990} from our results $\eta_{\rm{smoothed,D_2O}}$ (Table~\ref{Ragueneau_Smoothed}). The red curves show the uncertainty band ($\pm 1$ standard deviation) of our measurements.}
\end{figure}

Figure~\ref{Ragueneau_CompaLittD2O} compares the present data to previously published data sets~\cite{millero_density_1971,osipov_shear_1977,agayev_experimental_1980,agayev_heavy-water_1990}. Like in the case of light water, our data agree with Millero~\cite{millero_density_1971} and Agayev~\cite{agayev_experimental_1980,agayev_heavy-water_1990}, whereas they systematically deviate from Osipov's data~\cite{osipov_shear_1977}. The deviation is however less striking than in the case of light water, due to the lower precision of our D$_2$O data.

\begin{table*}
	\caption{\label{Ragueneau_TableauLitterature_H2O_Precis}Selected previous datasets for the viscosity of H$_2$O at or near atmospheric pressure}
		\begin{ruledtabular}
			\begin{tabular}{lcccc}
				First author and reference	&   Year		&			Accuracy ($\%$)						&		Temperature range (K)							&			Number of data			\\
				Hallet~\cite{hallett_temperature_1963}			&		1963		&			1													&			250.15 - 264.15										&			15									\\
				Korosi~\cite{korosi_viscosity_1968}			&		1968		&			0.3												&			348.15 - 423.15										&			7										\\
				Eicher~\cite{eicher_high-precision_1971}			&		1971		&			0.3												&			264.87 - 273.15										&			5										\\
  			Kestin~\cite{kestin_free_1981}			&		1981		&			0.3												&			343.70 - 423.96										&			4										\\
				Collings~\cite{collings_high_1983}	&		1983		&			0.2												&			274.15 - 288.15, 303.15 - 343.15	&			10									\\
				Kestin~\cite{kestin_viscosity_1985}			&		1985		&			0.5												&			458.55 - 491.95										&			3										\\
				Berstad~\cite{berstad_accurate_1988}		&		1988		&			0.05											&			293.15 - 298.15										&			6										\\
				Dehaoui~\cite{dehaoui_viscosity_2015}		&		2015		&			2.9												&			239.15														&			1										\\
			\end{tabular}
		\end{ruledtabular}
\end{table*}

\begin{table*}
	\caption{\label{RagueneauFitsPower}Best-fit parameters for a Speedy-Angell law (eq.~\ref{Ragueneau_PowerLaw}). The fit of $D_{s}$ and $\tau_{\theta}$ of light water is the same as previously published~\cite{dehaoui_viscosity_2015}.}
		\begin{ruledtabular}
			\begin{tabular}{lcccccc}
				Data $A$				    &	Temperature range (K)	&	Number of points $N$		&   $A_0$		&			$T_s$ (K) 					&		$\gamma$				&			Reduced $\chi_{2,r}=\chi_2/(N-3)$	\\
				$\eta$,	H$_2$O  &			239.15-348.15		&			49								&	$137.4 \pm 0.3~\rm{\mu Pa.s}$			&			$225.99\pm0.14$&	$1.636\pm0.004$	&	1.31							\\
				$D_{s}$,	H$_2$O		 &	237.8-498.2					&			36								&	$16077 \pm 78 \rm{\mu m^2/s}$	&	$213.96 \pm 0.35$		&	$-2.0801	\pm 0.0086$& 1.62	\\
				$\tau_{\theta}$, H$_2$O& 236.18-451.63        &			51								& $217.89\pm 0.90$ fs		  					&	$223.05 \pm 0.14$		&	$1.8760\pm 0.0065$& 0.61		\\			
				
				$\eta$, D$_2$O 		&			243.7-329.65	 &			70								&	$140.53 \pm 0.07~\rm{\mu Pa.s}$		&			$233.52 \pm 0.12$&	$1.599\pm0.003$	&	0.76							\\
				$D_{s}$, D$_2$O&			244.2-373.15		  &			28								&	$15720 \pm 140~\rm{\mu m^2/s}$ 	&			$220.1\pm 1.1$&	$-2.06\pm 0.03$	&	1.30							\\
				$\tau_{\theta}$, D$_2$O&			239.0-473.15		  &			47								&	$238.3\pm 1.5$ fs 	&			$230.79\pm 0.13$&	$1.811\pm 0.007$	&	0.77							
			\end{tabular}
		\end{ruledtabular}
\end{table*}

\section{Choice of literature data}
\label{RagueneauLiterature}

To test the SE and SED relations, we need to combine our viscosity data together with previously tabulated viscosity data in the stable liquid, and then with data for the self-diffusion coefficient and the rotational correlation time. In this section we explain how the literature data were chosen.

\subsection{Viscosity, self-diffusion and rotational correlation time of H$_2$O}

\begin{table*}
	\caption{\label{Ragueneau_TableauLitteratureD2O}Previous data sets for the viscosity of D$_2$O at or near atmospheric pressure.}
		\begin{ruledtabular}
			\begin{tabular}{lcccc}
				First author and reference		   &   year		  &			Accuracy 								&		Temperature range ($K$)			&			Number of data			\\
				Lewis~\cite{lewis_viscosity_1933}						 &		1933		&				$0.5\%$								&			$278.15$ to $308.15$			&					7								\\
				Taylor~\cite{taylor_properties_1934}					 &		1934		&				n.p.									&			$293.15$									&					1								\\
				Baker~\cite{baker_conductance_1935}						 &		1935		&				$0.1\%$								&			$298.15$									&					1								\\
				Jones~\cite{jones_viscosity_1936}						 &		1936		&				$0.01\%$							&			$298.15$									&					1							  \\
				Lemonde~\cite{lemonde_sur_1941}				 & 		1941		&				n.p.									&			$278.15$ to $290.15$			&					13						 \\
				Hardy~\cite{hardy_viscosity_1949}			 		   &		1949 		&				$0.1\%$								&			$278.15$ to $398.15$			&					11							\\
				Heiks~\cite{heiks_density_1954}						 &		1954		&		$3.10^{-6}$~Pa.s					&			$303.15$ to $523.15$			&					12						\\
				Harlow~\cite{harlow_further_1967}					 & 		1967		&				$1\%$									&			$283$ to $373$						&					7							\\
				Agaev~\cite{agaev_viscosity_1968}					 &		1968		&				$0.5\%$								&			$277.15$ to $348.15$			&					10							\\
				Selecki~\cite{selecki_viscosity_1970}			   &		1970		&				$0.8\%$								&			$298.15$ to $363.15$			&					6								\\
				Millero~\cite{millero_density_1971}  		   &		1971		&		$3$ to $8.10^{-7}$~Pa.s		&			$278.15$ to $343.15$			&					28		   				\\
				Kellom\"aki~\cite{kellomaki_viscosities_1975}  &		1975		&		    $0.1\%$						  	&			$283.15$ to $308.15$			&					6								\\
				Wilbur~\cite{wilbur_selfdiffusion_1976}					 & 		1976		&				$2\%$									&			$283$ to $363$						&					4							\\
				Jonas~\cite{jonas_molecular_1976}						 &		1976		&				$2\%$									&			$283$ to $363$						&					3							\\
				DeFries~\cite{defries_molecular_1977}				 &		1977		&				$2\%$									&			$278$ to $283$						&					2							\\
				Gon\c{c}alves~\cite{goncalves_viscosity_1980}&		1979 	  &		$0.02$ to $0.06\%$				&			$293.15$ to $333.15$			&					6							\\
				Agayev~\cite{agayev_experimental_1980}			   &		1980		&		$0.5$ to $1.2\%$					&			$277.01$ to $369.05$			&					12							\\
				Kestin~\cite{kestin_viscosity_1985}					 &		1985		&				$0.5\%$								&			$298.25$ to $493.05$			&					12							\\
				Agayev~\cite{agayev_heavy-water_1990}					 &		1989		&				$0.6\%$								&				$276.97$								&						1							\\
				Harris~\cite{harris_temperature_2004}					 &		2004		&  			$1\%$									&			$278.15$ to $298.15$			&					9							\\
			\end{tabular}
		\end{ruledtabular}
\end{table*}

\begin{table*}
	\caption{\label{Ragueneau_TableauLitterature_D2O_Precis}Selected previous data sets for the viscosity of D$_2$O at or near atmospheric pressure}
		\begin{ruledtabular}
			\begin{tabular}{lcccc}
				First author and reference&   year		&			Accuracy ($\%$)						&		Temperature range ($K$)							&			Number of data			\\
				Hardy~\cite{hardy_viscosity_1949}			&		1949		&				$0.1$										&			$353.15$ to $398.15$							&					7 							\\
				Selecki~\cite{selecki_viscosity_1970}	&		1970		&				$0.8\%$									&			$348.15$													&					1								\\				
				Millero~\cite{millero_density_1971} &		1971		&		$3$ to $8.10^{-7}$~Pa.s	  	&			$278.15$ to $308.15$,							&					22 		   				\\
																	&						&											 				  &			$318.15$, $328.15$, 							& 												\\
																	&						&																&			$338.15$, $343.15$								&													\\
				Gon\c{c}alves~\cite{goncalves_viscosity_1980}&1979&		$0.03$ to $0.05\%$  				&			$313.15$ to $333.15$							&					3								\\
				Agayev ~\cite{agayev_experimental_1980}	&		1980		&		$0.5$ to $1.2\%$						&			$277.01$ to $300.75$							&					11							\\
																	&						&																&			$305.59$ to $369.05$							&													\\
				Kestin~\cite{kestin_viscosity_1985}		&		1985		&				$0.5\%$									&			$298.25$ to $493.05$							&					12							\\
				Agayev~\cite{agayev_heavy-water_1990}	  &		1989		&				$0.6\%$						  		&				$276.97$										 		&						1							\\		
			\end{tabular}
		\end{ruledtabular}
\end{table*}

\begin{table*}
	\caption{\label{Ragueneau_TableauLitteratureD2O_D}Data sets for the self-diffusion coefficient of D$_2$O at or near atmospheric pressure.}
		\begin{ruledtabular}
			\begin{tabular}{lcccc}
				First author and reference				  &   year		&			Accuracy 					&		Temperature range ($K$)			&			Number of data			\\
				Longsworth~\cite{longsworth_mutual_1960}			&		1960		&			$0.1\%$						&				278.15-318.15						&					3								\\
				Devell~\cite{devell_measurements_1962}							&		1962		&			$1\%$							&				298.15									&					1								\\
				Murday~\cite{murday_selfdiffusion_1970}							&		1970		&			$7\%$							&				294.3-295.2							&					3								\\
				Mills~\cite{mills_self-diffusion_1973}								&		1973		&			$0.2\%$						&				278-318									&					3								\\
				Wilbur~\cite{wilbur_selfdiffusion_1976}							&		1976		&			$10\%$						&				283-363									&					4								\\
				Woolf~\cite{woolf_tracer_1976}								&		1976		&			$2\%$							&				280.5-328.2							&					7								\\
				DeFries~\cite{defries_molecular_1977}						&		1977		&			n.p.							&				278-283									&					2								\\
				Weing\"artner~\cite{weingartner_diffusion_1984}	&		1984		&			$0.4$ to $0.6\%$	&				281-318									&					3								\\		
				Prielmeier~\cite{prielmeier_pressure_1988}			&		1988		&			$6\%$							&				258.0-363.0							&					11							\\
				Price~\cite{price_temperature_2000}								&		2000		&			$5\%$							&				244.2-298.25						&					17							\\
				Hardy~\cite{hardy_isotope_2001}								&		2001		&			$1.4$ to $1.9\%$	&				288-328									&					7								\\
				Yoshida\footnotemark[1]\footnotetext{Pressure up to $167\,\mathrm{bar}$.}~\cite{yoshida_new_2005}						&		2005		&			$0.5$ to $3\%$		&				303-623									&					18							\\
				Yoshida\footnotemark[1]~\cite{yoshida_self-diffusion_2008}						&		2008		&			$1$ to $10\%$			&				303-623									&					33							\\
			\end{tabular}
		\end{ruledtabular}
\end{table*}

\begin{table*}
	\caption{\label{Ragueneau_TableauLitterature_D2O_Precis_D}Selected data sets for the self-diffusion coefficient of D$_2$O at or near atmospheric pressure.}
		\begin{ruledtabular}
			\begin{tabular}{lcccc}
				First author and reference				  &   year		&			Accuracy 					&		Temperature range ($K$)			&			Number of data			\\
				Mills~\cite{mills_self-diffusion_1973}								&		1973		&			$0.2\%$						&				278-318									&					3								\\
				Woolf~\cite{woolf_tracer_1976}								&		1976		&			$2\%$							&				280.5, 313.2						&					3								\\
				Weing\"artner~\cite{weingartner_diffusion_1984}	&		1984		&			$0.6\%$						&				281											&					1								\\		
				Prielmeier~\cite{prielmeier_pressure_1988}			&		1988		&			$6\%$							&				332.5-363.0							&					2 							\\
				Price~\cite{price_temperature_2000}								&		2000		&			$5\%$							&				244.2-276.40						&					13							\\
				Hardy~\cite{hardy_isotope_2001}								&		2001		&			$1.4$ to $1.9\%$	&	288, 293, 303, 308, 328				&					5								\\
				Yoshida\footnote{Pressure up to $167\,\mathrm{bar}$.}~\cite{yoshida_self-diffusion_2008}						&		2008		&			$1$ to $10\%$			&				373-623									&					26							\\
			\end{tabular}
		\end{ruledtabular}
\end{table*}

The most accurate sources of viscosity data for H$_2$O are gathered in Table~\ref{Ragueneau_TableauLitterature_H2O_Precis}. We used Eicher's data as re-evaluated by Kestin~\cite{kestin_viscosity_1978}. Eicher's, Colling's, Korosi's and Kestin's (1985) data were recalculated using the current reference value for viscosity of pure light water at $293.15$~K, $\eta(T_0)=1.0016\,\mathrm{mPa\,s}$~\cite{huber_new_2009}. Kestin's and Korosi's data are measured under pressure, but below $20$~bars. The weak pressure dependence of dynamic parameters allows neglecting the difference between those data and the atmospheric pressure value. Berstad's data are calculated using the provided fits from 20 to $25^\circ$C every $1^\circ$C.

To generate SE and SED plots for H$_2$O, we combine individual viscosity data points with Speedy-Angell fits to self-diffusion and rotational correlation time:
\begin{equation}
	A(T)=A_0\left(\frac{T}{T_s}-1\right)^{-\gamma} \, .
	\label{Ragueneau_PowerLaw}
\end{equation}
The best-fit parameters for $D_{s}$ and $\tau_{\theta}$ were given previously~\cite{dehaoui_viscosity_2015}; for easy reference, they are recalled in Table~\ref{RagueneauFitsPower}.

\begin{table*}
	\caption{\label{Table_D_Woolf_Weingartner}Self-diffusion of D$_2$O deduced from the measurements of Woolf~\cite{woolf_tracer_1976} and Weing\"artner~\cite{weingartner_diffusion_1984}. For each $D_{s}$ value the right reference is given.}
	\begin{ruledtabular}
		\begin{tabular}{lcccccccccc}
			$T$~(K)		&		280.5	&	280.5	&	281.2	&	288.2	&	288.2	&	298.2	&	298.2	&	313.2	&	318.2	&	328.2\\
			$D_{s}~\rm{\times 10^9 m^2/s}$	&		1.05~\cite{woolf_tracer_1976}	&	1.06~\cite{woolf_tracer_1976}	&	1.12~\cite{weingartner_diffusion_1984}&	1.36~\cite{woolf_tracer_1976}	&	1.37~\cite{woolf_tracer_1976}	&	1.87~\cite{woolf_tracer_1976}	& 1.87~\cite{weingartner_diffusion_1984}	&	2.67~\cite{woolf_tracer_1976}	&	2.99~\cite{weingartner_diffusion_1984}&	3.60~\cite{woolf_tracer_1976}
		\end{tabular}
	\end{ruledtabular}
\end{table*}

\subsection{Viscosity, self-diffusion and rotational correlation time of D$_2$O}

For D$_2$O, we carefully collected all literature data that we could find and chose the most accurate values at each available temperature. We checked that at common temperatures, the selected data were in good agreement with each other. 

The uncertainties given in the sources were considered as $1-\sigma$ uncertainties (confidence interval $68\%$), unless otherwise stated in the article. Most authors provide uncertainties without stating whether the temperature uncertainty was propagated in the uncertainty on $\eta$, $D_{s}$ or $\tau_{\theta}$. It is obviously not the case in Millero~\cite{millero_density_1971}, Gon\c{c}alves~\cite{goncalves_viscosity_1980}, Matubayasi~\cite{matubayasi_structural_2001} and Hardy~\cite{hardy_isotope_2001} since propagating the temperature uncertainty results in an uncertainty on $\eta$, $D_{s}$ or $\tau_{\theta}$ greater than the uncertainty provided in their papers. As a consequence, fitting their data by a Speedy-Angell law (eq.~\ref{Ragueneau_PowerLaw}) and propagating the temperature uncertainty, we obtained the accuracies indicated in the tables. For other papers, the uncertainty on the temperature is compatible with the uncertainty on $\eta$, $D_{s}$ or $\tau_{\theta}$ and we decided to keep the provided uncertainty.

Due to the scant and often confusing information provided by the authors about their uncertainties, some differences exist between our tables~\ref{Ragueneau_TableauLitteratureD2O} and~\ref{Ragueneau_TableauLitteratureD2O_D} and the ones provided by Assael~\cite{assael_new_2021} and Su\'arez-Iglesias~\cite{suarez-iglesias_self-diffusion_2015}.

\subsubsection{Viscosity of $D_2O$}

The existing sources of viscosity data for D$_2$O are gathered in Table~\ref{Ragueneau_TableauLitteratureD2O}. Following Matsunaga~\cite{matsunaga_transport_1983}, we discarded the data of Heiks~\cite{heiks_density_1954}. Following Kestin~\cite{kestin_viscosity_1985}, we discarded the data provided in 1968 by Agaev and Yusibova~\cite{agaev_viscosity_1968}.

The most accurate sources we decided to use are gathered in Table~\ref{Ragueneau_TableauLitterature_D2O_Precis}. Note that we recalculated the data of Hardy, Millero, Kestin and Gon\c{c}alves using the current reference value for viscosity of pure light water at $293.15$~K, $\eta(T_0)=1.0016\,\mathrm{mPa\,s}$~\cite{huber_new_2009}. We chose to discard Jones' data point since it disagrees with Millero's data point at the same temperature, while Millero agrees with all the other authors at any temperature. We decided to consider only the data up to $T=500\,\mathrm{K}$. At that temperature, the vapor pressure is less than $30$ bars, and dynamic properties of heavy water are nearly constant in that pressure range.

\subsubsection{Self-diffusion of $D_2O$\label{sec:DD2O}}

The existing sources of self-diffusion data for D$_2$O are gathered in Table~\ref{Ragueneau_TableauLitteratureD2O_D}. The data of Longsworth~\cite{longsworth_mutual_1960} were later reanalyzed by Mills~\cite{mills_self-diffusion_1973}, and we used the latter set of data. 

Following Yoshida~\cite{yoshida_self-diffusion_2008} and Su\'arez-Iglesias~\cite{suarez-iglesias_self-diffusion_2015}, we discarded the data published by Yoshida in 2005~\cite{yoshida_new_2005}. Care must be taken about the data of Yoshida~\cite{yoshida_self-diffusion_2008}, that were measured under vapor pressure, because this reaches up to $167\,\mathrm{bar}$ at the highest temperature.

The data scatter of Price~\textit{et al.}~\cite{price_self-diffusion_1999} is not compatible with the uncertainties they provide, as already stated about their data on H$_2$O~\cite{dehaoui_viscosity_2015}. From the data scatter around a Speedy-Angell law fit, we estimate the actual uncertainty to $5\%$.

DeFries~\cite{defries_molecular_1977} does not provide any uncertainty in their article. We chose to apply the value provided by Wilbur~\cite{wilbur_selfdiffusion_1976}, since they come from the same group, on the same equipment, at a one year interval.

The data of Woolf~\cite{woolf_tracer_1976} and Weing\"artner~\cite{weingartner_diffusion_1984} were deduced from their measurements of the diffusion coefficient of DTO in D$_2$O in the same way as Mills~\cite{mills_self-diffusion_1973} deduced the self-diffusion of H$_2$O and D$_2$O from the diffusion of isotopes~\cite{mills_self-diffusion_1973,longsworth_mutual_1960}. The results are reproduced in the table~\ref{Table_D_Woolf_Weingartner}.

The most accurate sources we decided to use are gathered in Table~\ref{Ragueneau_TableauLitteratureD2O_Precis_T1}.

The ratios $D_{s,\mathrm{H_2O}} / D_{s,\mathrm{D_2O}}$ of Hardy~\cite{hardy_isotope_2001} were graphically read on their figure 2 and $D_{s,\mathrm{D_2O}}$ was deduced using the value of $D_{s,\mathrm{H_2O}}$ calculated following Eq.~\ref{Ragueneau_PowerLaw} with parameters given in Table~~\ref{RagueneauFitsPower}. An uncertainty on the reading was added to those data. 

The most accurate sources we decided to use are gathered in Table~\ref{Ragueneau_TableauLitterature_D2O_Precis_D}.

\subsubsection{Rotational correlation time of $D_2O$}

\begin{table*}
	\caption{\label{Ragueneau_TableauLitteratureD2O_T1}Data sets for the rotational correlation time $\tau_{\theta}$ of D$_2$O at or near atmospheric pressure.}
		\begin{ruledtabular}
			\begin{tabular}{lcccc}
				First author and reference				  &   year		&			Accuracy ($\%$)			&		Temperature range ($K$)			&			Number of data		\\
				Hindman~\cite{hindman_relaxation_1971}						&		1971		&			3										&		263	- 447										&			24								\\
				Hindman~\cite{hindman_relaxation_1973}						&		1973		&			3 to 10$\%$					&		236.1	-	287.6								&			35								\\
				Jonas~\cite{jonas_molecular_1976}								&		1976		&			$3\%$								&		283 - 363										&			3									\\
				DeFries~\cite{defries_molecular_1977}						&		1977		&			$3\%$								&		278 - 283										&			2									\\
				Lang~\cite{lang_pressure_1980}									&		1980		&			$10\%$							&		239 - 283										&			15								\\	
				Matubayasi~\cite{matubayasi_structural_2001}			&		2001		&			$3$ to $13\%$				&		303 - 613										&			8									\\
				Ropp~\cite{ropp_rotational_2001}									&		2001		&			$1\%$								&		275.7 - 310.5								&			8									\\
				Hardy~\cite{hardy_isotope_2001}								&		2001		&			$2$ to $3\%$				&		278 - 358										&			17								\\
				Qvist~\cite{qvist_rotational_2012}								&		2012		&			$0.5$ to $2\%$			&		241 - 307										&			21								\\
			\end{tabular}
		\end{ruledtabular}
\end{table*}

\begin{table*}
	\caption{\label{Ragueneau_TableauLitteratureD2O_Precis_T1}Selected data sets for the rotational correlation time $\tau_{\theta}$ of D$_2$O at or near atmospheric pressure.}
		\begin{ruledtabular}
			\begin{tabular}{lcccc}
				First author and reference				  &   year		&			Accuracy 					&		Temperature range (K)				&			Number of data			\\
				Hindman~\cite{hindman_relaxation_1971}						&		1971		&			3										&		367	- 447										&			11								\\
				Jonas~\cite{jonas_molecular_1976}								&		1976		&			$3\%$								&		363													&			1								\\
				Lang~\cite{lang_pressure_1980}									&		1980		&			$10\%$							&		239													&			1								\\	
				Matubayasi~\cite{matubayasi_structural_2001}			&		2001		&			$5\%$								&		473													&			1									\\
				Ropp~\cite{ropp_rotational_2001}									&		2001		&			$1\%$								&		310.5												&			1									\\
				Hardy~\cite{hardy_isotope_2001}								&		2001		&			$2$ to $3\%$				&		308 - 358										&			11								\\
				Qvist~\cite{qvist_rotational_2012}								&		2012		&			$0.5$ to $2\%$			&		241 - 307										&			21								\\
			\end{tabular}
		\end{ruledtabular}
\end{table*}

\begin{table*}
	\caption{\label{RagueneauFitsVTF}Best-fit parameters for a VTF law (eq.~\ref{RagueneauVTFLaw})}
		\begin{ruledtabular}
			\begin{tabular}{lcccccc}
				Data				  &	Temperature range (K)	&	Number of points $N$		&   $A_0$		&			$B$ (K) 					&		$T_0$ (K)				&			Reduced $\chi_{2,r}=\chi_2/(N-3)$			\\
				$\eta$,	H$_2$O 		&			239.15-348.15			  &			49								&	$43.3 \pm 0.4~\rm{\mu Pa.s}$	&			$394\pm 3$&	$167.6\pm0.4$	&	18							\\
				$\eta$, D$_2$O 		&			243.7-329.65					  &			70								&	$46.1 \pm 0.3~\rm{\mu Pa.s}$		&			$402.6 \pm 1.6$&	$171.1\pm0.3$	&	6.8							\\
				$D_{s}$, D$_2$O&			244.2-373.15		  &			28								&	$75000 \pm 5000~\rm{\mu m^2/s}$ 	&			$-510\pm 20$&	$159\pm 3$	&	1.53							\\
				$\tau_{\theta}$, D$_2$O&			239.0-473.15		  &			47								&	$87.5 \pm 1.5$ fs 	&			$361\pm 3$&	$184.9\pm 0.4$	&	28							
			\end{tabular}
		\end{ruledtabular}
\end{table*}

The existing sources of rotational correlation time data for D$_2$O are gathered in Table~\ref{Ragueneau_TableauLitteratureD2O_T1}. Most of the authors provide values of the spin-lattice relaxation time $T_1$. Qvist~\cite{qvist_rotational_2012} provides the rotational correlation time $\tau_{\theta}=\frac{1}{\omega_Q^2T_1}$, $\omega_Q=1.02.10^6\rm{s^{-1}}$ being the nuclear quadrupole frequency of deuterium. Hardy~\textit{et al.}~\cite{hardy_isotope_2001} provide the ratio between the apparent rotational diffusion coefficients $D_\mathrm{rot}=1/(6\,\tau_{\theta})$ of light and heavy water. We decided to convert all those data into the rotational correlation time $\tau_{\theta}$.

The data of Hindman~\cite{hindman_relaxation_1971} and Matubayasi~\cite{matubayasi_structural_2001} are measured on the liquid-vapor equilibrium line. We decided to keep all the data of Hindman since in that temperature range, the vapor pressure does not exceed $10$ bars, while the rotational correlation time is nearly constant in that pressure range. For the same reasons as above, we used Matubayasi's data only up to $500$~K.

Hindman~\cite{hindman_relaxation_1971} and Qvist~\cite{qvist_rotational_2012} do not provide their raw data. We read the temperature where they measured the rotationnal correlation time $\tau_{\theta}$ (Qvist) or the spin-lattice relaxation time $T_1$ (Hindman) on their figures and computed the expected value from the fits they provide. Ropp's data~\cite{ropp_rotational_2001} are directly read on their Fig.~3. A reading uncertainty was added. Hardy's data~\cite{hardy_isotope_2001} are read as explained in Section~\ref{sec:DD2O} to deduce the rotational correlation time of D$_2$O from the fit of the rotational correlation time of H$_2$O (Eq.~\ref{Ragueneau_PowerLaw} with parameters given in Table~\ref{RagueneauFitsPower}).

\section{Discussion}
\label{Ragueneau_Discussion}
\subsection{Classic fits}

\begin{figure*}
	\includegraphics[width=12.9cm]{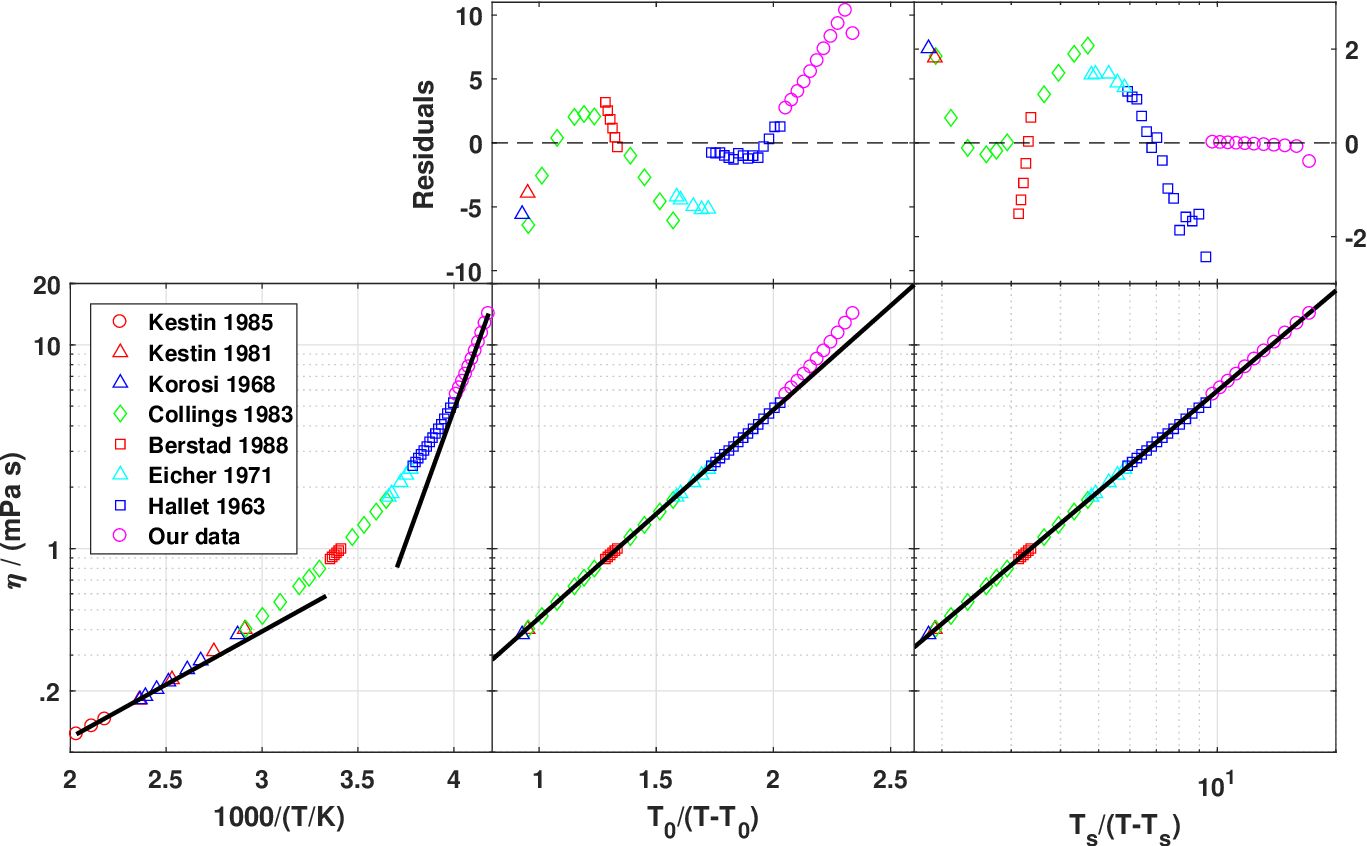}
	\caption{Viscosity of light water in different representations. The three panels
include the 8 datasets tabulated in Table~~\ref{Ragueneau_TableauLitterature_H2O_Precis} and our data (pink circles), together with a best-fit with various models (lines, see below). The lowest temperature pink circle is the one that was previously published in Ref.~\cite{dehaoui_viscosity_2015}. (Left) Arrhenius plot, showing an apparent activation energy increasing from 1200 to 6000 K upon cooling (solid lines). (Center) VTF representation, with best-fit parameters in Table~\ref{RagueneauFitsVTF}). (Right) Speedy-Angell representation, with best-fit parameters in Table~\ref{RagueneauFitsPower}).
(Top) The reduced residuals $\left(\eta_\mathrm{exp} - \eta_\mathrm{fit}\right)/\sigma_\mathrm{exp}$, where
$\eta_\mathrm{exp}$ and $\eta_\mathrm{fit}$ are the experimental and fitted viscosity, respectively, and $\sigma_\mathrm{exp}$ is the experimental uncertainty (1 SD). Note the different vertical scale in the top right panel.}
	\label{Ragueneau_FitsClassiques_H2O}
\end{figure*}

\begin{figure*}
	\includegraphics[width=12.9cm]{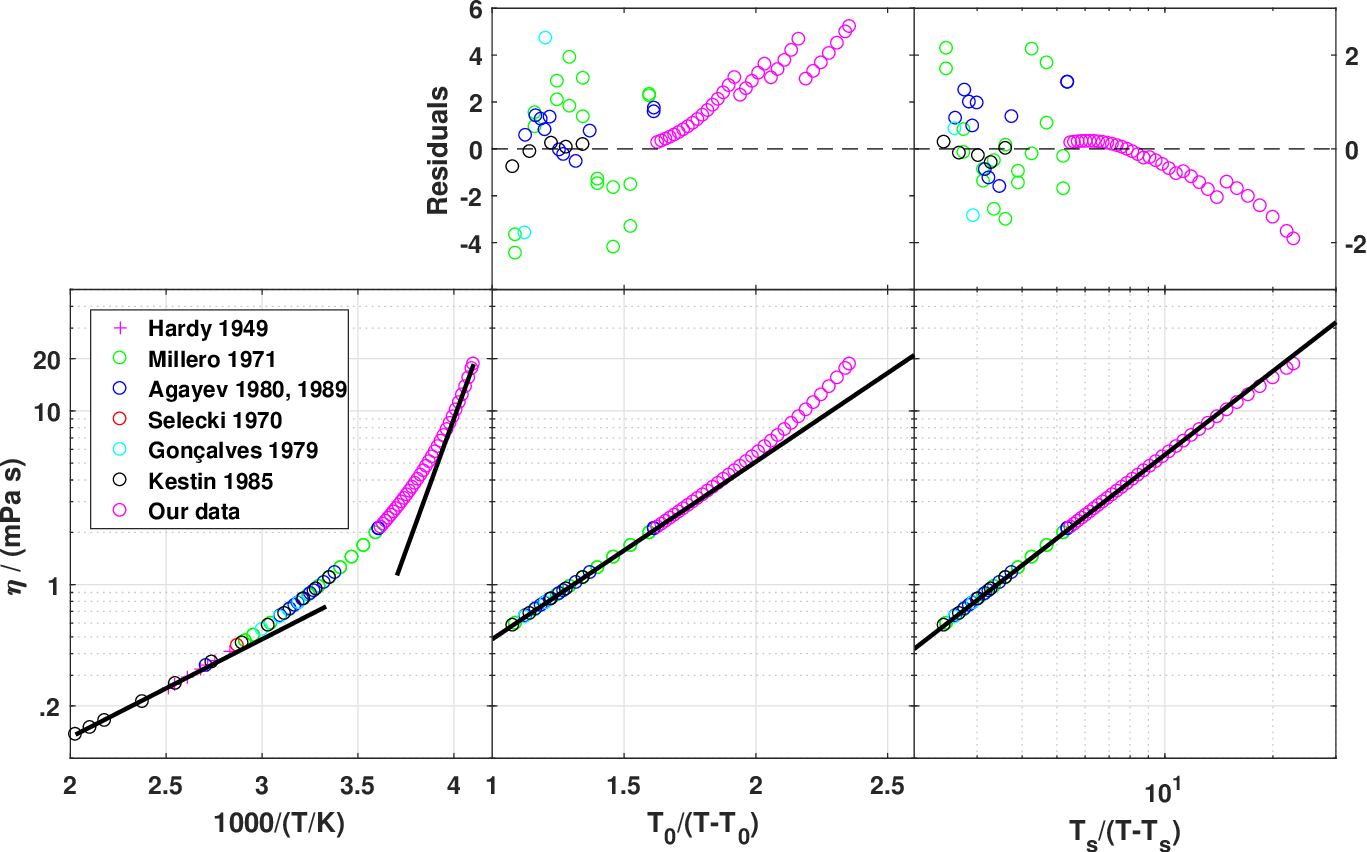}
	\caption{Viscosity of heavy water in different representations. The three panels
include the 7 datasets tabulated in Table~~\ref{Ragueneau_TableauLitterature_D2O_Precis} and our data (pink circles), together with a best-fit with various models (lines, see below). (Left) Arrhenius plot, showing an apparent activation energy increasing from 1300 to 7000 K upon cooling (solid lines). (Center) VTF representation, with best-fit parameters in Table~\ref{RagueneauFitsVTF}). (Right) Speedy-Angell representation, with best-fit parameters in Table~\ref{RagueneauFitsPower}).
(Top) The reduced residuals $\left(\eta_\mathrm{exp} - \eta_\mathrm{fit}\right)/\sigma_{\rm{exp}}$, where
$\eta_\mathrm{exp}$ and $\eta_\mathrm{fit}$ are the experimental and fitted viscosity, respectively, and $\sigma_\mathrm{exp}$ is the experimental uncertainty (1 SD). Note the different vertical scale in the top right panel.}
	\label{Ragueneau_FitsClassiques_D2O}
\end{figure*}

The viscosity of light and heavy water clearly departs from an Arrhenius behaviour, as shown in Figs.~\ref{Ragueneau_FitsClassiques_H2O} and~\ref{Ragueneau_FitsClassiques_D2O} (left panels), showing that both isotopes behave as fragile glassformers.

The viscosity of H$_2$O and D$_2$O and the diffusion coefficient and rotational correlation time of D$_2$O were fitted by the classic expressions: the Speedy-Angell power-law (Eq.~\ref{eq:etaSpeedyAngell} or Eq.~\ref{Ragueneau_PowerLaw}), and the Vogel-Tamman-Fulcher (VTF) law:
\begin{equation}
A(T)=A_0 \exp\left(\frac{B}{T-T_0}\right)\,.
\label{RagueneauVTFLaw}
\end{equation}

The temperature ranges and best-fit parameters are given in Tables~\ref{RagueneauFitsPower} and~\ref{RagueneauFitsVTF}. The fits of viscosity are presented in Figs.~\ref{Ragueneau_FitsClassiques_H2O} and~\ref{Ragueneau_FitsClassiques_D2O} (center and right panels).

\begin{figure}
	\includegraphics[width=8.6cm]{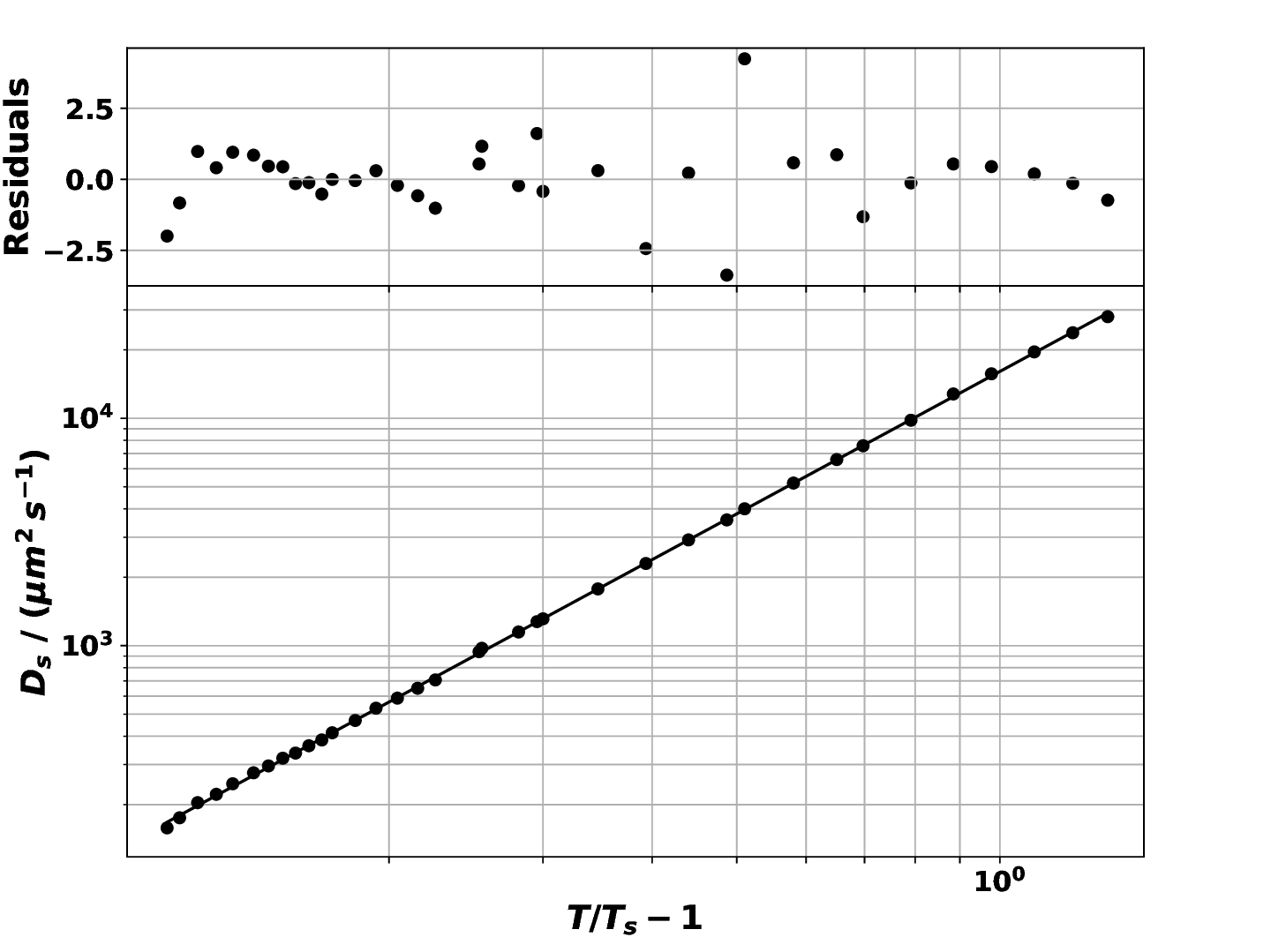}
	\caption{Fit of the self-diffusion coefficient of light water by a Speedy-Angell law, with parameters given in Table~\ref{RagueneauFitsPower}.}
	\label{Ragueneau_FitD_H2O}
\end{figure}

\begin{figure}
	\includegraphics[width=8.6cm]{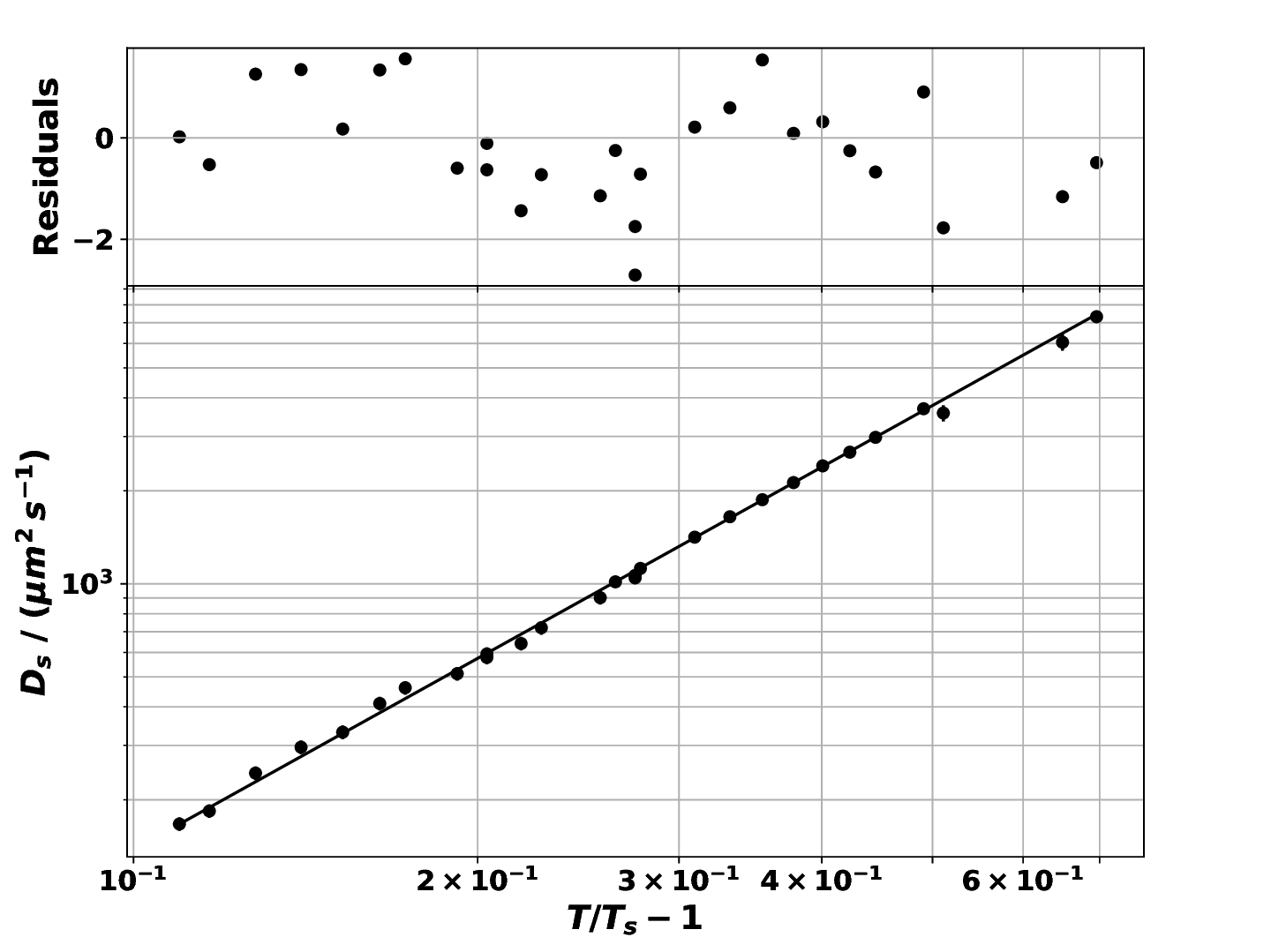}
	\caption{Fit of the self-diffusion coefficient of heavy water by a Speedy-Angell law, with parameters given in Table~\ref{RagueneauFitsPower}.}
	\label{Ragueneau_FitD_D2O}
\end{figure}

\begin{figure}
	\includegraphics[width=8.6cm]{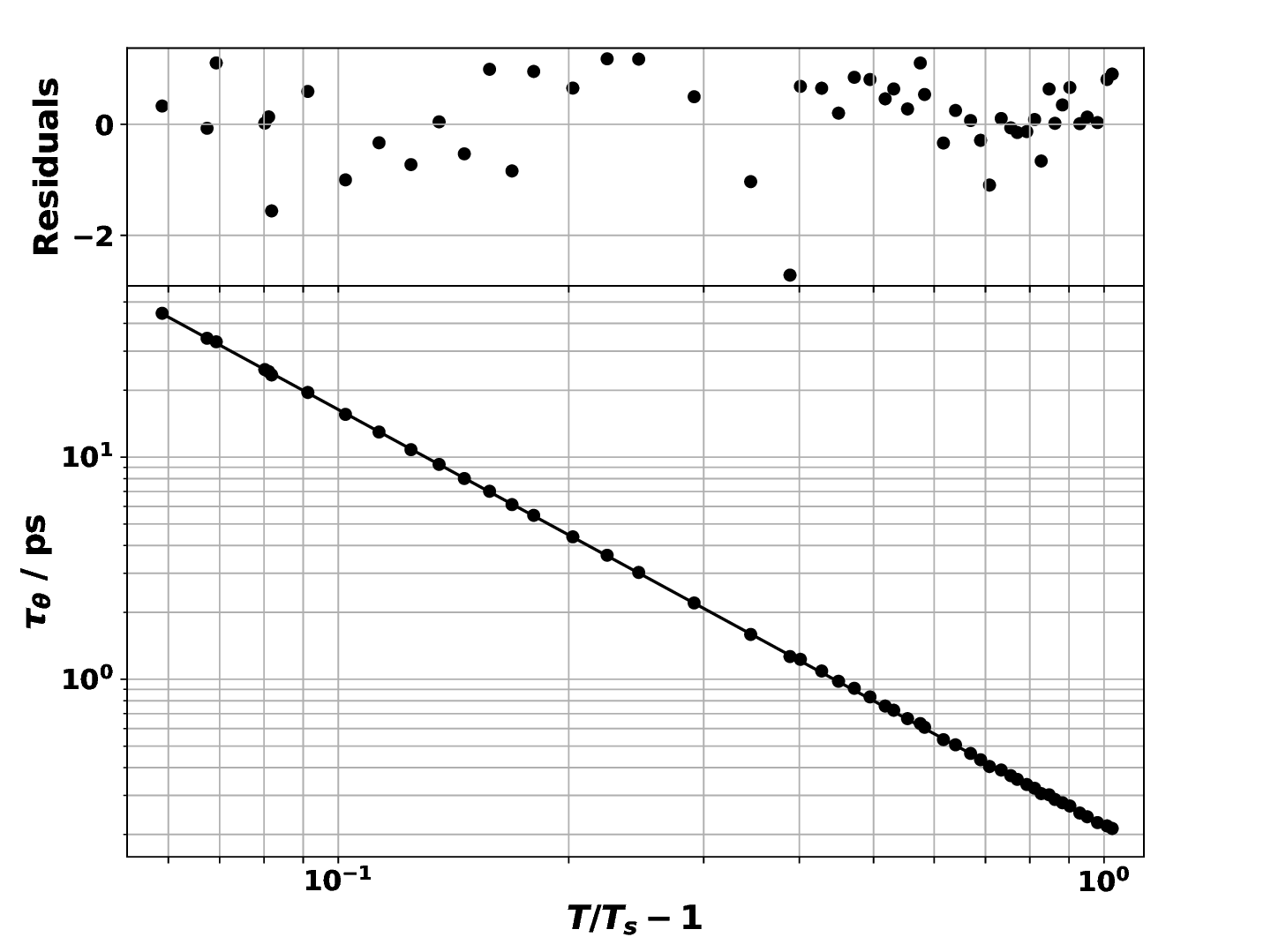}
	\caption{Fit of the rotational correlation time of light water by a Speedy-Angell law, with parameters given in Table~\ref{RagueneauFitsPower}.}
	\label{Ragueneau_Fittaur_H2O}
\end{figure}

\begin{figure}
	\includegraphics[width=8.6cm]{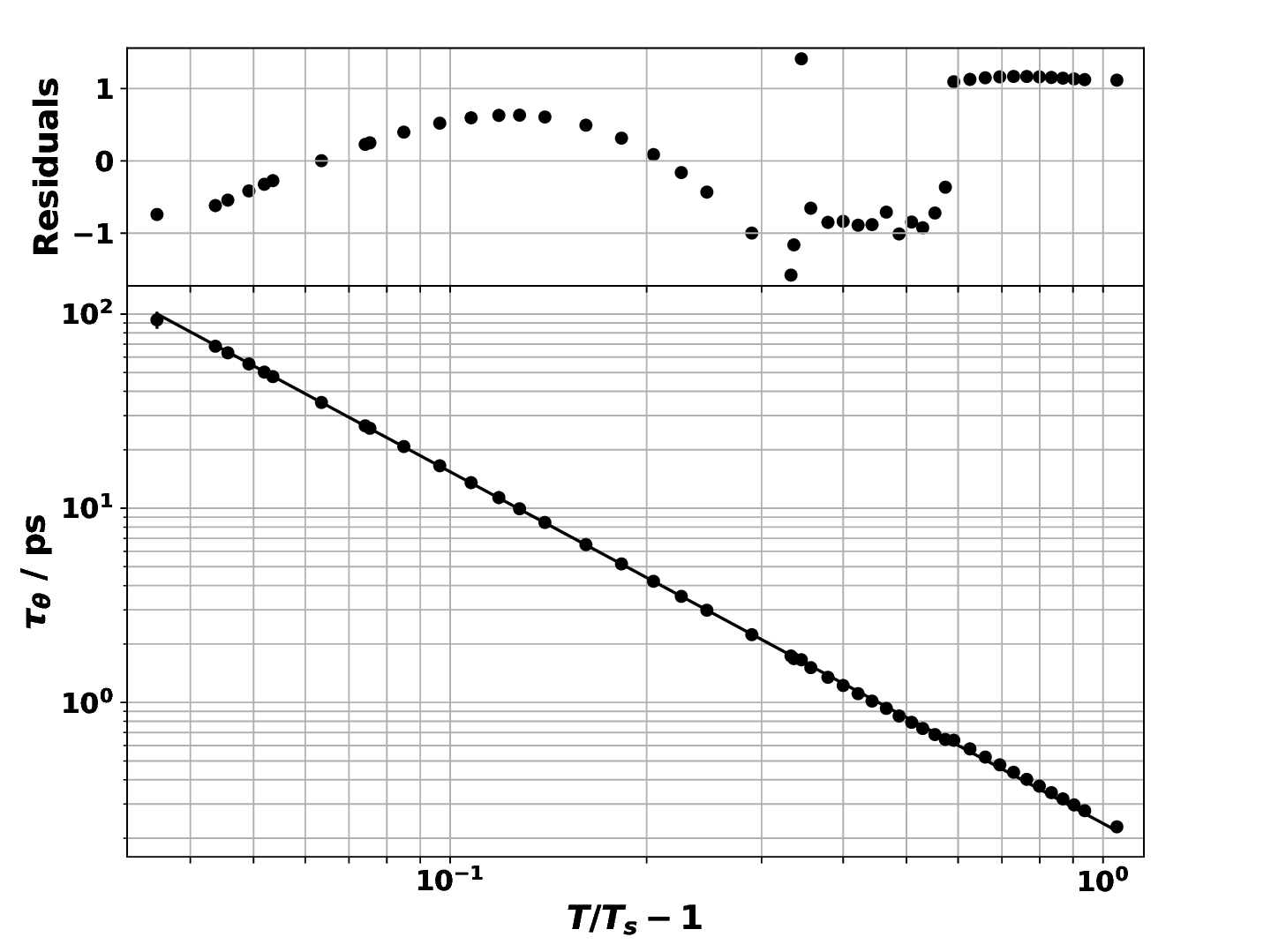}
	\caption{Fit of the rotational correlation time of heavy water by a Speedy-Angell law, with parameters given in Table~\ref{RagueneauFitsPower}.}
	\label{Ragueneau_Fittaur_D2O}
\end{figure}

As expected from previous results~\cite{dehaoui_viscosity_2015}, the Speedy-Angell law best describes dynamic data of light and heavy water over a large temperature range. The rotational correlation time is even well described by a Speedy-Angell law up to $473.15$~K, even if the data are taken on the saturation line at the highest temperatures. The fits of $D_{s}$ and $\tau_{\theta}$ by a Speedy-Angell law are shown in Fig.~\ref{Ragueneau_FitD_H2O} to~\ref{Ragueneau_Fittaur_D2O}.

\subsection{Comparison with mode coupling theory}

\begin{figure}
	\includegraphics[width=8.6cm]{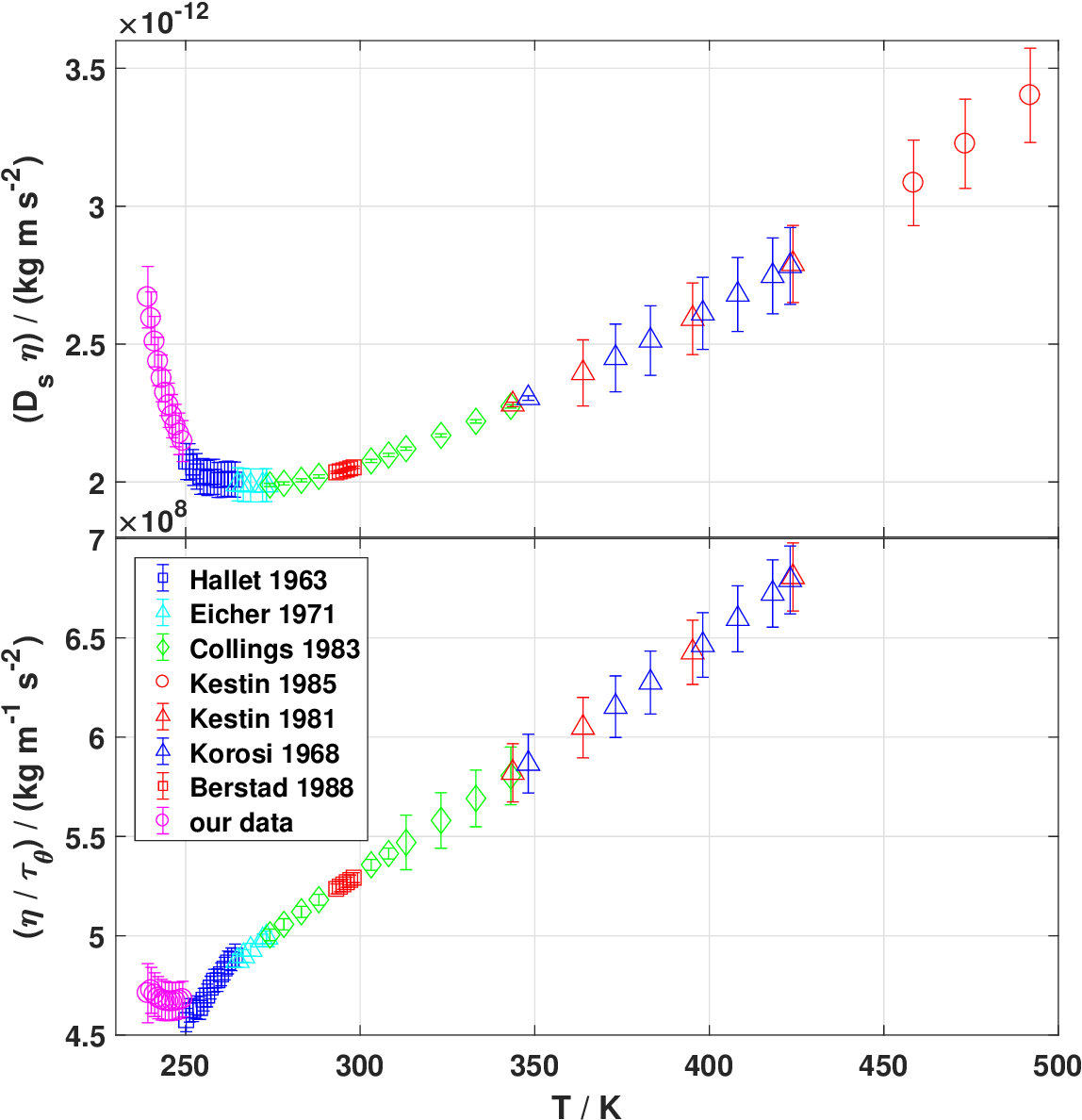}
	\caption{Test of the predictions of the mode coupling theory on the dynamic quantities of light water: $D_{s}\eta$ (top) and $\eta/\tau_{\theta}$ (bottom) should be constant according to the mode coupling theory.}
	\label{Ragueneau_TCM_H2O}
\end{figure}

\begin{figure}
	\includegraphics[width=8.6cm]{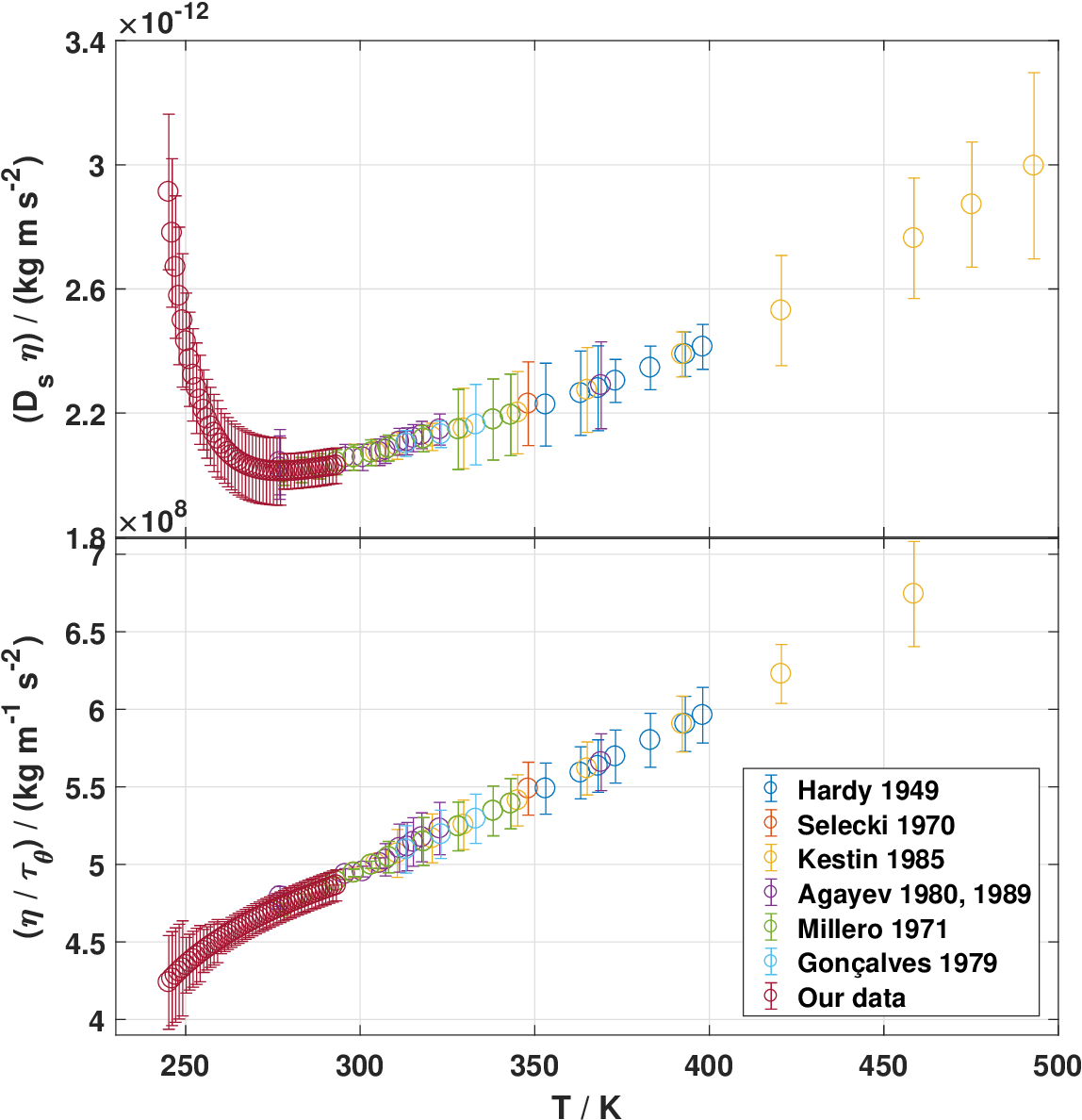}
	\caption{Test of the predictions of the mode coupling theory on the dynamic quantities of heavy water: $D_{s}\eta$ (top) and $\eta/\tau_{\theta}$ (bottom) should be constant according to the mode coupling theory.}
	\label{Ragueneau_TCM_D2O}
\end{figure}

Mode coupling theory~\cite{gotze_relaxation_1992} predicts a power-law dependence of dynamic quantities, similar to the Speedy-Angell formula. However, in the case of mode coupling, $T_{s}$ refers to the mode-coupling temperature, and should be the same for all three quantities ($D_s$, $\eta$, and $\tau_{\theta}$). This is not what is experimentally observed, neither for light nor for heavy water. Mode coupling theory also predicts the products $D_{s}\eta$ and $\eta/\tau_{\theta}$ to be temperature independent. Indeed, $D_{s}\eta$ is constant over a small temperature range ($\approx 260$ to $300$~K) for both isotopes, as can be seen in Figs.~\ref{Ragueneau_TCM_H2O} and~\ref{Ragueneau_TCM_D2O}, but it is not the case of $\eta/\tau_{\theta}$. Moreover, $D_{s}\eta$ increases sharply at lower temperatures, in contradiction with the predictions of mode coupling theory. Mode coupling theory is usually expected to be valid close enough to the glass transition temperature, a condition not met here. Still, experiments on $\rm{Zr_{64}Ni_{36}}$ show that mode-coupling could hold up to $2T_s$~\cite{brillo_relation_2011}. This is not the case for water: mode coupling theory fails in describing the dynamic quantities of light and heavy water in the experimentally accessible temperature range.

\begin{table}
	\caption{\label{ParamsFitDHT}Best-fit parameters for the function $\ln[D_{s}/\mathrm{(m\,s^{-2})}]=\sum_{i=0}^5 A_i\left(\ln(T/\mathrm{K})\right)^i$ from 244 to 623~K. $\chi_2=2.23$}
		\begin{ruledtabular}
			\begin{tabular}{cc}
				$i$				  &	$A_i$				\\
					0					&		$-1.655461201959378.10^5$		\\
					1					&		$1.365289265726531.10^5$	\\
					2					&		$-4.506424270204171.10^4$	\\
					3					&		$7.439738201650022.10^3$		\\
					4					&		$-6.142825158988378.10^2$		\\
					5					&		$20.292383150993103$
			\end{tabular}
		\end{ruledtabular}
\end{table}

\subsection{Stokes-Einstein and Stokes-Einstein-Debye relations}

\begin{figure}
	\includegraphics[width=8.6cm]{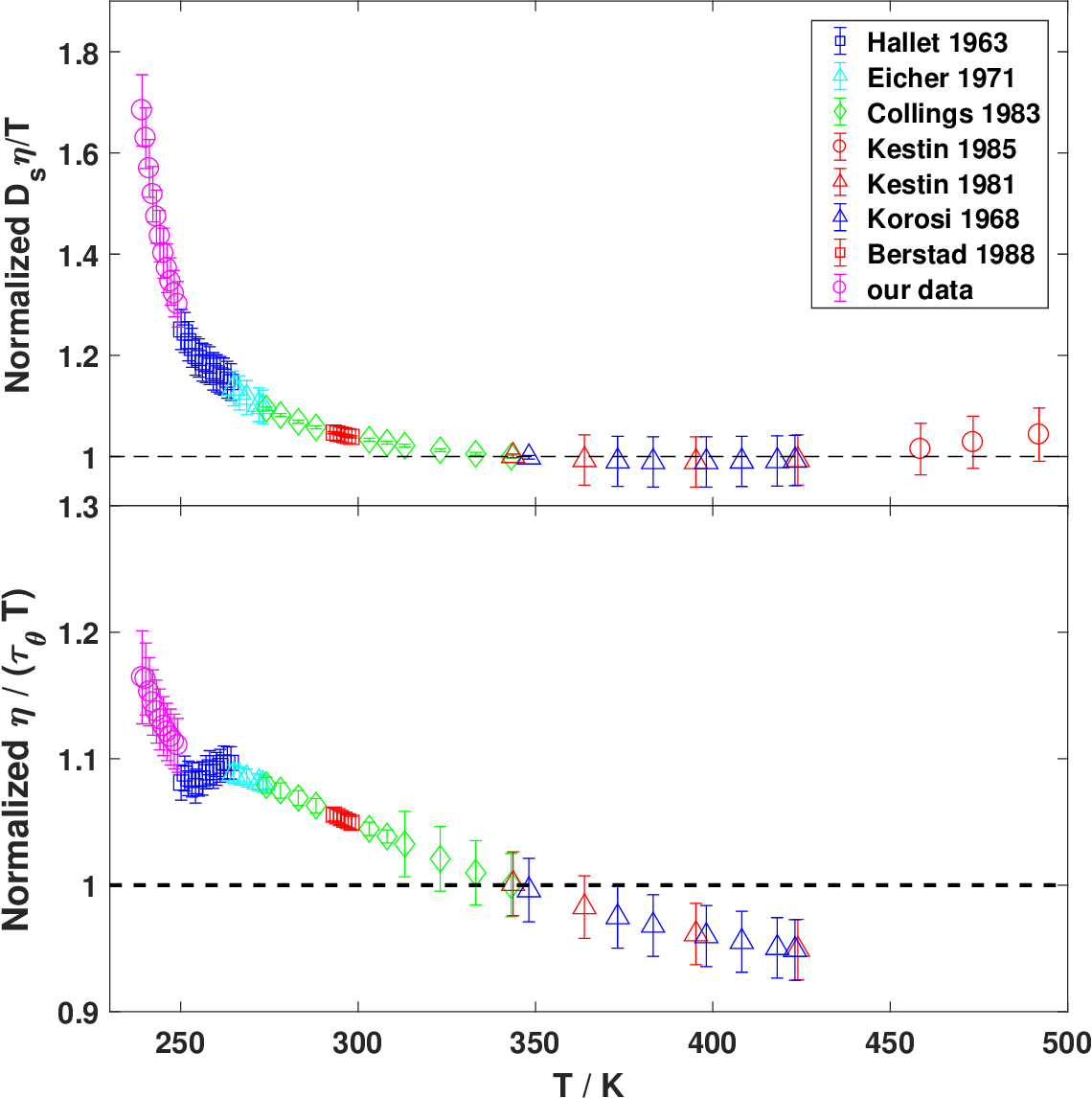}
	\caption{\label{Ragueneau_StokesEinstein_H2O}Stokes-Einstein (top) and Stokes-Einstein-Debye (bottom) ratios as a function of temperature in light water.}
\end{figure}

\begin{figure}
	\includegraphics[width=8.6cm]{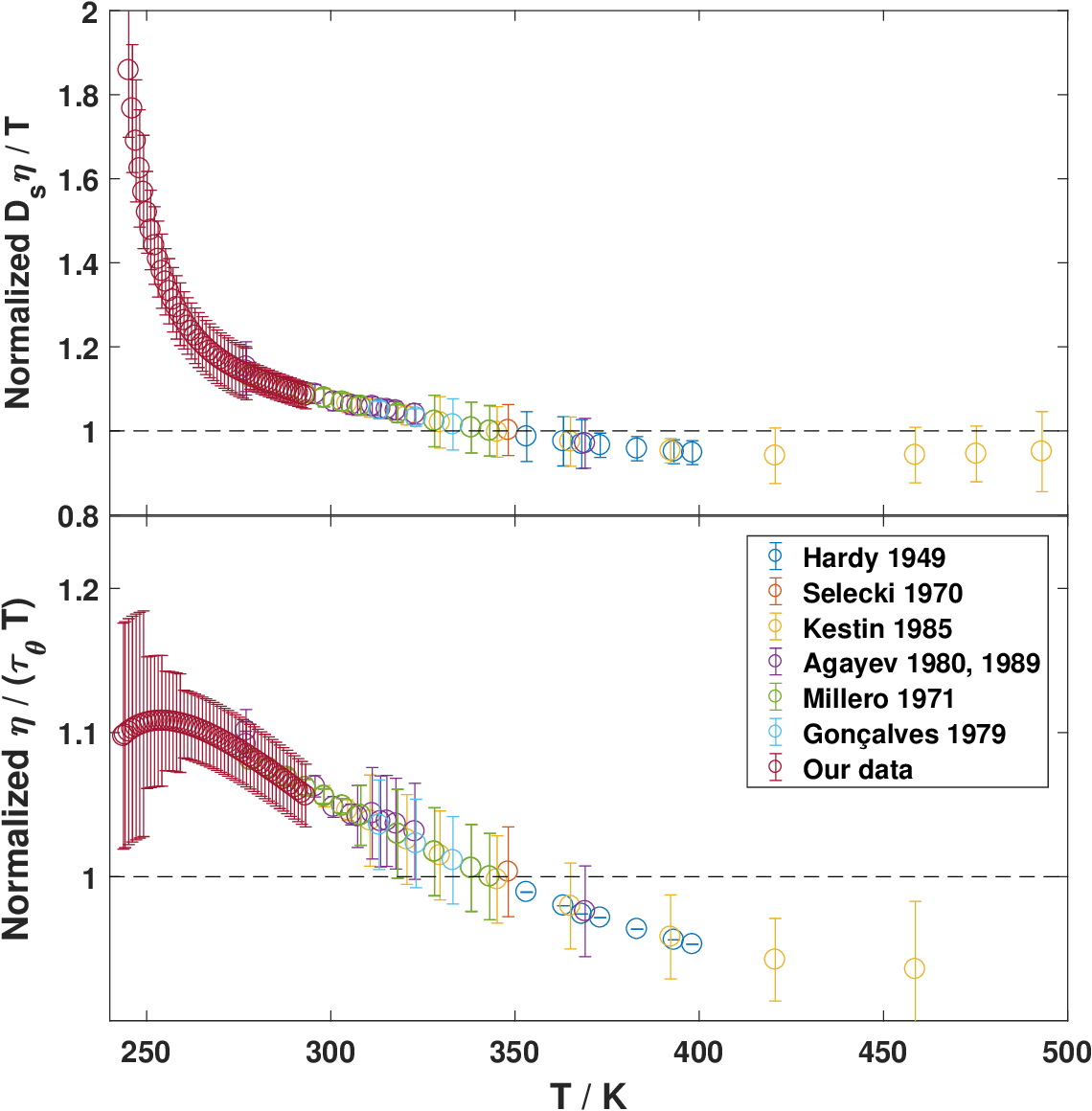}
	\caption{\label{Ragueneau_StokesEinstein_D2O} Stokes-Einstein (top) and Stokes-Einstein-Debye (bottom) ratios as a function of temperature in heavy water.}
\end{figure}

Using the selected data for the viscosity $\eta$, self-diffusion $D_{s}$ and rotational correlation time $\tau_{\theta}$ of heavy water (Tables~\ref{Ragueneau_TableauLitterature_D2O_Precis},~\ref{Ragueneau_TableauLitterature_D2O_Precis_D},~\ref{Ragueneau_TableauLitteratureD2O_Precis_T1}), we plot as a function of temperature in Figs.~\ref{Ragueneau_StokesEinstein_H2O} and~\ref{Ragueneau_StokesEinstein_D2O} the SE ratio $D_{s}\eta / T$ (upper panels) and the SED ratio $\eta / (T\tau_{\theta})$ (lower panels). In all panels the data are normalized by the value at $343.15\,\mathrm{K}$. To generate all graphs presented in this section, we used the data points for $\eta$ provided by the different authors, together with values of $D_{s}$ and $\tau_{\theta}$ computed at the same temperatures by a fitting function. For $\tau_{\theta}$, we use the Speedy-Angell fit with parameters given in Table~\ref{RagueneauFitsPower}. For $D_{s}$, as the Speedy-Angell fit fails at very high temperature, we use from 244 to $623\,\mathrm{K}$ the following function: $\ln[D_{s}/ \mathrm{(m\,s^{-2})} ]=\sum_{i=0}^5 A_i \left(\ln(T/ \mathrm{K} )\right)^i$, with best-fit parameters $A_i$'s given in Table~\ref{ParamsFitDHT}.

The results are quite similar in light and heavy water: SE and SED relations both hold at high temperature, but are increasingly violated as temperature decreases. The SE violation reaches $70\%$ to $90\%$ at the lowest temperature, with an increasing trend as the temperature decreases, while the SED violation only reaches $15\%$ to $20\%$. Because the failure of SE and SED is progressive, it is not possible to precisely define a temperature at which the violation starts. Still, we can note that the degree of violation becomes significant around the triple points ($273.16\,\mathrm{K}$ for H$_2$O and $276.969\,\mathrm{K}$ for D$_2$O), where all ratios reach around 1.1. In usual fragile glassformers, SE violation starts around $1.2\,T_g$~\cite{chang_heterogeneity_1997}, where $T_{g}$ is the glass transition temperature. For water, $T_{g}$ is reported to be 131.8 and $135.6\,\mathrm{K}$ for H$_2$O and D$_2$O, respectively~\cite{shephard_molecular_2016}. Therefore, SE violation already starts at anomalously high temperatures, above $2T_{g}$. It is important to note that this violation of the SE relation applies to the self-diffusion of water molecules, not to the Brownian diffusion of colloids that we used to deduce the viscosity of the liquids (Eq.~\ref{StokesEinstein}). Indeed, the sphere diameter we use ($350\,\mathrm{nm}$) is sufficiently large for hydrodynamic laws to hold.

We refer the reader to Ref.~\cite{caupin_giant_2022} for a discussion of the possible origin of the SE violation in water. In particular, it is shown there that the SE ratio for D$_2$O are equal to those for H$_2$O, shifted by $+7\,\mathrm{K}$ in temperature. This suggests that the decoupling in both isotopes, albeit starting at different temperatures, may have a common origin, possibly related to the putative liquid-liquid transition in supercooled water~\cite{gallo_water_2016}.

The SE ratio in light water is in good agreement with the results from molecular dynamic simulations of the viscosity and the self-diffusion of light water at ambient pressure using the TIP4P/2005 water model~\cite{montero_de_hijes_viscosity_2018,dueby_decoupling_2019} and the TIP4P/2005f model~\cite{guillaud_decoupling_2017}. In the latter study, the authors emphasize that the viscosity and the $\alpha$-relaxation time are strongly coupled above $309\,\mathrm{K}$ but get uncoupled below that temperature. When investigating the SE and SED relations below that temperature, the shear-viscosity needs to be computed instead of the $\alpha$-relaxation time as was frequently done before~\cite{becker_fractional_2006,kumar_relation_2007}. 

The $\alpha$-relaxation time has already been determined experimentally twice using the Optical Kerr Effect~\cite{torre_structural_2004, taschin_evidence_2013}. In both studies it is found to be well described by a Speedy-Angell law with different fit parameters: in 2004 the authors determined $T_{s}=221\pm5\,\mathrm{K}$ and $\gamma=2.2\pm 0.3$. In 2013 they found $T_s=227\,\mathrm{K}$ and $\gamma=1.7$ on a larger temperature range without providing their uncertainty. The precision on those experimental values is too low to allow concluding experimentally about the decoupling between the viscosity and the $\alpha$-relaxation time.

Turning to the SED relation, the violation remains moderate down to $250\,\mathrm{K}$, which is in line with simulations using the TIP4P/2005 water model~\cite{kawasaki_spurious_2019}; see however Section I of the Supplemental Material for a more detailed discussion of SED relation in water.

\subsection{Apparent hydrodynamic radius}

\begin{figure}
	\includegraphics[width=8.6cm]{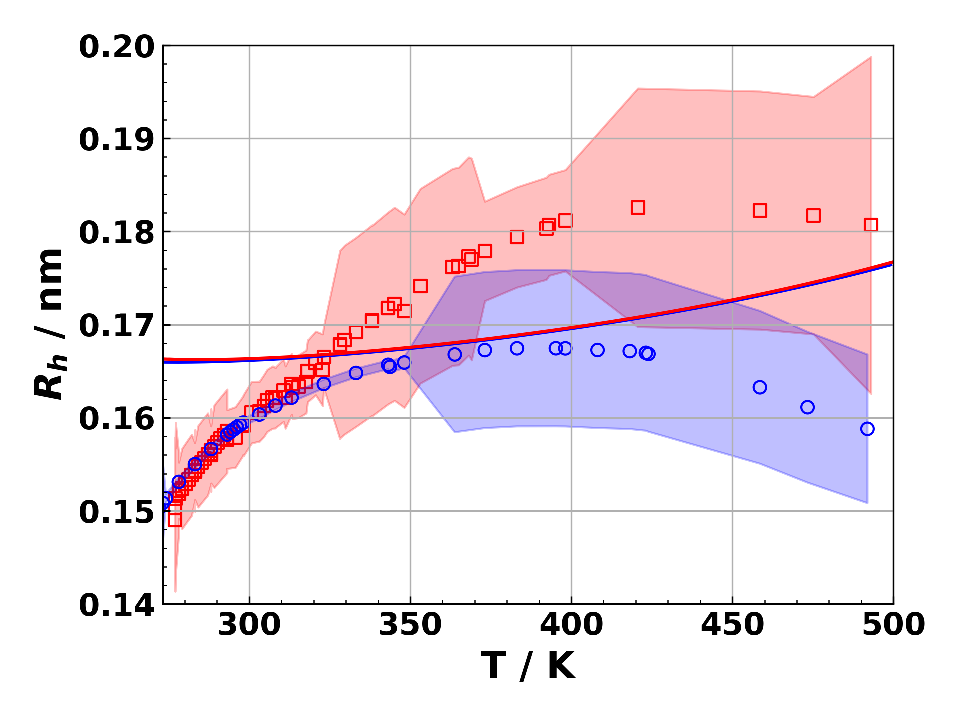}
	\caption{\label{Ragueneau_RayonHydro}Hydrodynamic radius $R_h$ as a function of temperature calculated from Eq.~\ref{Eq_HydroRadius} for H$_2$O (blue circles)
and D$_2$O (red squares); the colored areas indicate the $1-\sigma$ uncertainty. The solid curves show $R_\mathrm{rcp}$ from Eq.~\ref{Eq_Rrcp} for H$_2$O (blue) and D$_2$O (red).}
\end{figure}

The SE ratios displayed in Figs.~\ref{Ragueneau_StokesEinstein_H2O} and~\ref{Ragueneau_StokesEinstein_D2O} are normalized. This hides quantitative information about the effective hydrodynamic radius $R_h$ of a water molecule, which can be defined by analogy with the Stokes-Einstein equation for a Brownian sphere, Eq.~\ref{StokesEinstein}:
\begin{equation}
	R_{h}=\frac{k_BT}{C\pi\eta D_{s}} \, ,
	\label{Eq_HydroRadius}
\end{equation}
$C$ being a coefficient varying from $4$ to $6$ when the condition at the surface of the sphere changes from a full-slip to a no-slip condition, respectively. Choosing $C=4$ yields $R_h$ displayed in Fig.~\ref{Ragueneau_RayonHydro}. 

The definition of $R_h$ with Eq.~\ref{Eq_HydroRadius} is meaningful only when the SE relation holds; therefore, we plot only values above the triple point. Both isotopes have close values of $R_h\simeq 0.17\,\mathrm{nm}$. The $10\%$ difference at the highest temperatures is not significant, as there is some uncertainty in the diffusion coefficient values at high temperatures, with up to $10\%$ discrepancy between authors for H$_2$O at $373\,\mathrm{K}$~\cite{yoshida_self-diffusion_2008, krynicki_pressure_1978}. It is interesting to compare $R_h$ to a typical molecular size. We define it from the molar volumes $V_\mathrm{mol}$ of H$_2$O~\cite{wagner_international_1993} and D$_2$O~\cite{herrig_reference_2018}, as the radius $R_\mathrm{rcp}$ of hard spheres whose random close packing leads to the same molar volume: 
\begin{equation}
R_\mathrm{rcp} = \left(\frac{3}{4\pi}\,
\frac{\phi \,V_\mathrm{mol}}{\mathcal{N}_A} \right)^{1/3} \,.
\label{Eq_Rrcp}
\end{equation}
Here $\phi=0.64$ is the compaction for random-close packing, and $\mathcal{N}_A$ the Avogradro constant. Figure~\ref{Ragueneau_RayonHydro} shows that $R_\mathrm{rcp}$ and $R_{h}$ are in good agreement for both isotopes above $300\,\mathrm{K}$. This is a rather surprising result, as the continuous medium approximation behind the SE relation would be expected to fail at the molecular level.

\begin{figure}
	\includegraphics[width=8.6cm]{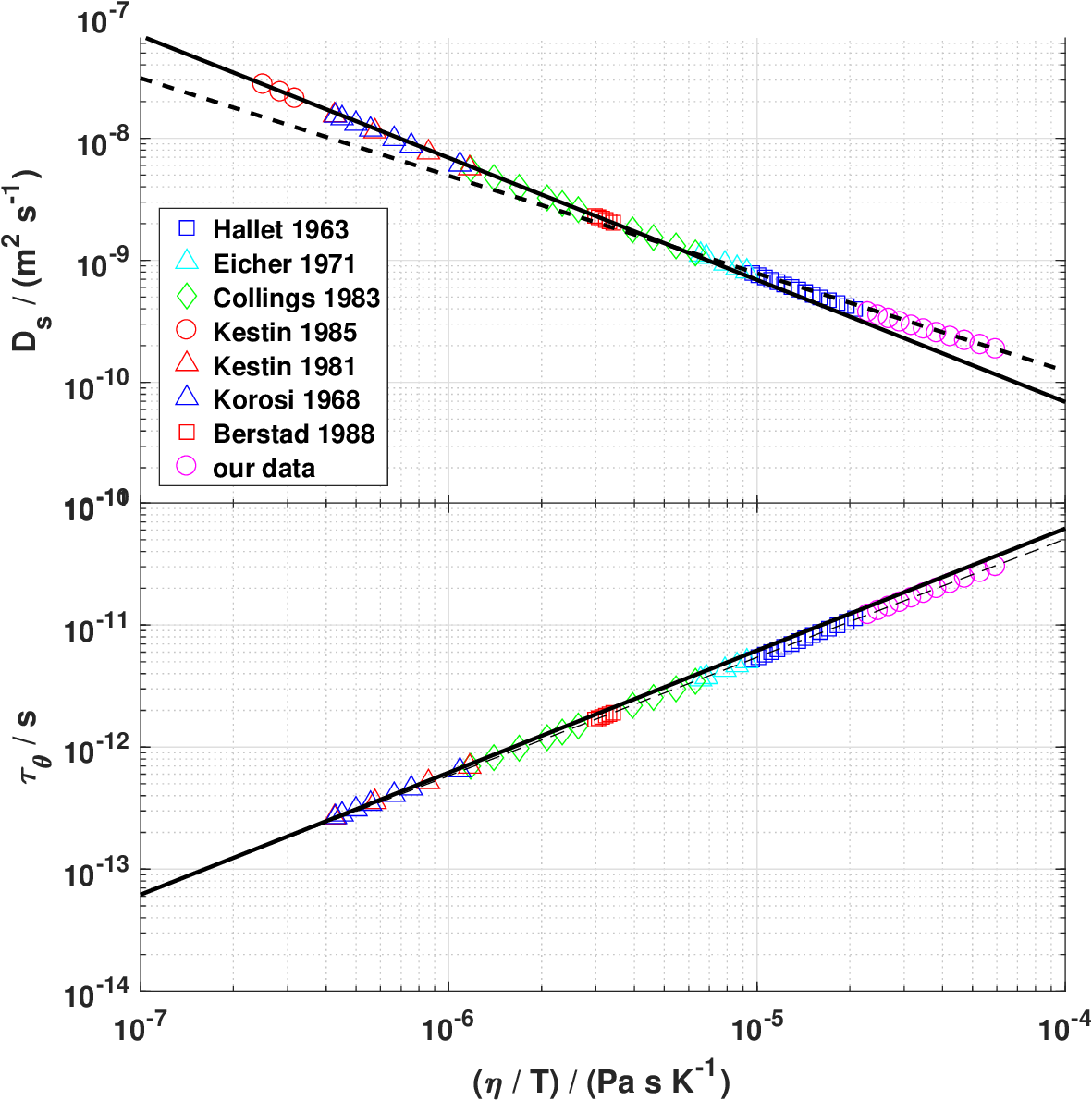}
	\caption{\label{Ragueneau_StokesEinsteinFractionnaire_H2O}Fractional Stokes-Einstein (top) and Stokes-Einstein-Debye (bottom) ratios as a function of viscosity in light water. Straight lines: Stokes-Einstein and Stokes-Einstein-Debye relations. Dashed line: $D_{s}\propto\left(\eta/T\right)^{-0.8}$ (top) ; $\tau_{\theta}\propto\left(\eta/T\right)^{0.9704}$ (bottom).}
\end{figure}

\begin{figure}
	\includegraphics[width=8.6cm]{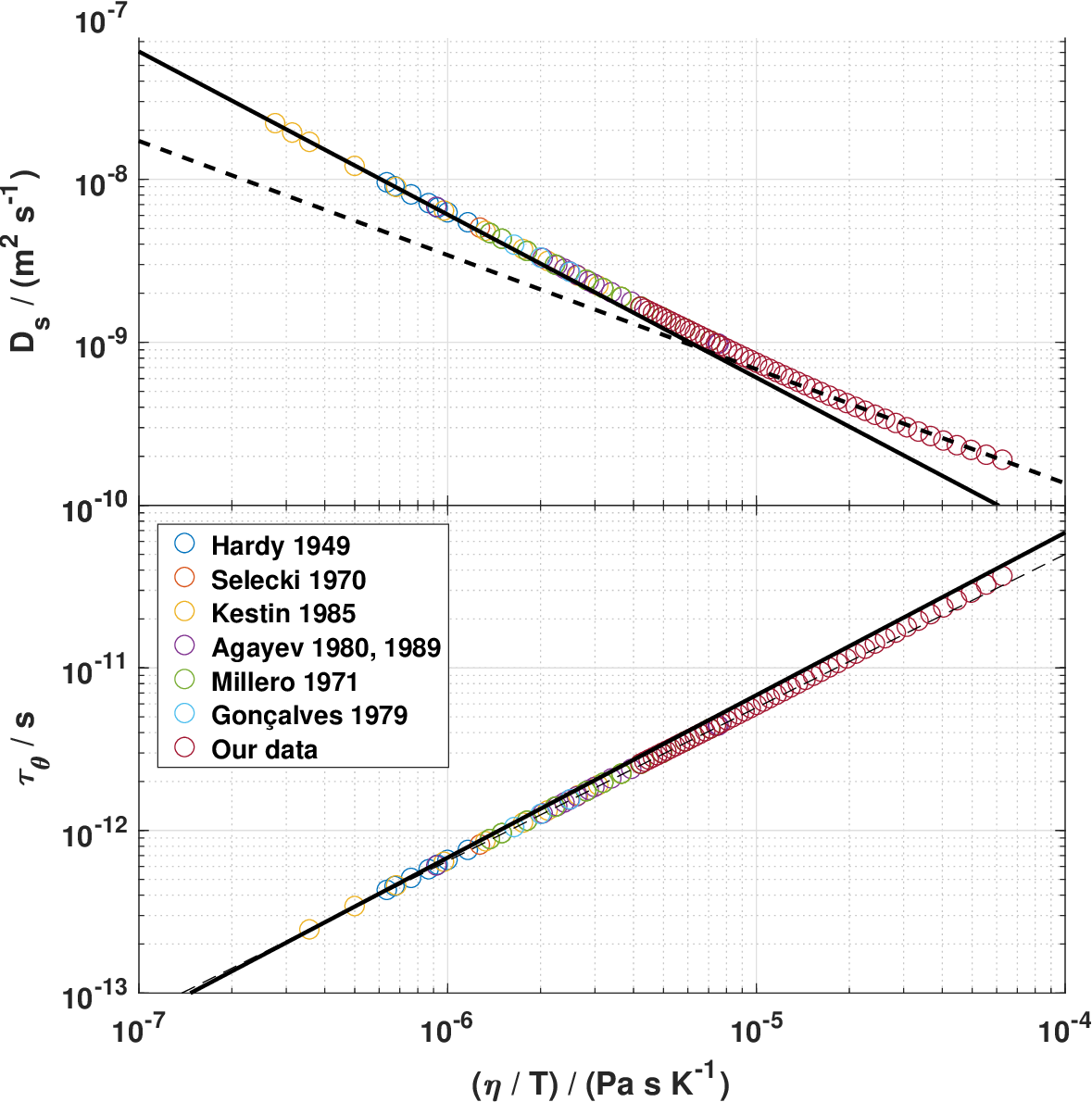}
	\caption{\label{Ragueneau_StokesEinsteinFractionnaire_D2O}Fractional Stokes-Einstein (top) and Stokes-Einstein-Debye (bottom) ratios as a function of viscosity in heavy water. Straight lines: Stokes-Einstein and Stokes-Einstein-Debye relations. Dashed line: $D_{s}\propto\left(\eta/T\right)^{-0.7}$ (top) ; $\tau_{\theta}\propto\left(\eta/T\right)^{0.9567}$ (bottom).}
\end{figure}

\subsection{Fractional Stokes-Einstein and Stokes-Einstein-Debye relations}

To account for the deviations from the SE relation, the so-called fractional SE relations have been introduced~:
\begin{equation}
\frac{D_{s}}{T} \propto \eta ^ {-t} \, ,
\label{eq:FSE}
\end{equation}
or alternatively:
\begin{equation}
D_{s} \propto \left( \frac{ \eta }{ T } \right) ^ {-\zeta} \, .
\label{eq:FSE-T}
\end{equation}
Such relations perform well for a variety of liquids, as well as for simulation data on the Lennard-Jones fluid (see Ref.~\cite{harris_fractional_2009} for a review). To justify the fractional SE relations, some theoretical arguments have been given~\cite{harris_fractional_2009}. An obstruction model~\cite{douglas_obstruction_1998}, which assumes diffusion amid obstructions representing dynamic heterogeneities, predicts Eq.~\ref{eq:FSE-T} with $\zeta=3/5$. An entropic barrier hopping theory of glassy hard sphere colloidal suspensions~\cite{schweizer_activated_2004} finds fractional SE behavior with $\zeta$ ranging from 0.73 to 0.90, depending on the density fluctuation correlation length. For the rotational correlation time $\tau_\theta$, a fractional SED relation similar to Eq.~\ref{eq:FSE-T},
\begin{equation}
\tau_\theta \propto \left( \frac{ \eta }{ T } \right)^\zeta \, ,
\label{eq:FSED-T}
\end{equation}
has been first used for a tracer in ortho-terphenyl~\cite{andreozzi_study_1996}.

We test Eqs.~\ref{eq:FSE-T} and~\ref{eq:FSED-T} in Figs.~\ref{Ragueneau_StokesEinsteinFractionnaire_H2O} and~\ref{Ragueneau_StokesEinsteinFractionnaire_D2O} for light and heavy water, respectively. At high temperatures, the SE relation holds ($\zeta = 1$). At low temperatures, $\zeta$ switches from $1$ to $0.8$ (resp. $0.7$) in H$_2$O (resp. D$_2$O). The fractional Stokes-Einstein-Debye relation Eq.~\ref{eq:FSED-T} holds with a constant exponent 0.9704 in H$_2$O and 0.9567 in D$_2$O on the whole temperature range for both isotopes. The experimental behavior is similar to that found in simulations. In a study of the ST2 water model~\cite{becker_fractional_2006}, the $\alpha$-relaxation time $\tau_\alpha$ was used as a proxy for viscosity. $D_s$ and $\tau_\theta$ were analyzed with the analogs of Eqs.~\ref{eq:FSE-T} and~\ref{eq:FSED-T}, respectively. Upon cooling, $\zeta$ for $D_s$  switches from $1$ to $0.7-0.8$, while $\zeta=0.9$ represents well $\tau_\theta$ at low temperature. In Ref.~\cite{kawasaki_spurious_2019}, $\eta$ was directly simulated instead of $\tau_\alpha$ and suggests $\zeta$ switching from $1$ to $\simeq 0.7$ for $D_s$ and $\zeta \gtrsim 0.8$ for $\tau_\theta$.

\section{Conclusion\label{sec:conclusion}}

We have obtained new measurements of the viscosity of H$_2$O with a better accuracy than in our previous work~\cite{dehaoui_viscosity_2015}, and the first reliable measurements of the viscosity of deeply supercooled D$_2$O. It appears that previous measurements on heavy water by Osipov~\cite{osipov_shear_1977} are biased, possibly due to an electroosmosis effect, as we already suggested in the case of light water~\cite{dehaoui_viscosity_2015}. The temperature dependence of dynamic properties of D$_2$O (viscosity, self-diffusion, and rotational correlation time) obeys a Speedy-Angell power-law, like in the case of H$_2$O. The violation of the SE relation is similar in both isotopes, reaching at the lowest temperature above 70\% and 80\% for light and heavy water, respectively. The SED relation also shows a similar behavior in H$_2$O and D$_2$O, with only a mild violation less than 20\%. Even if those two features are characteristic of fragile glassformers close to the glass transition temperature, as predicted by mode coupling theory, both isotopes significantly differ from other glassformers. The main difference is the unusually high temperature at which decoupling between dynamic quantities occurs, more than twice the glass transition temperature, while this decoupling occurs around $1.2T_g$ in usual glassformers. Deviations from the predictions of the mode coupling theory are also observed.

The decoupling between dynamic quantities in molecular glassformers is often attributed to dynamic heterogeneities~\cite{ediger_spatially_2000}. Our results suggest that dynamic heterogeneities exist in both light and heavy water far above the glass transition temperature. Many molecular dynamic simulations find such heterogeneities and SE violation on light water~\cite{becker_fractional_2006,mazza_connection_2007,kumar_relation_2007,limmer_putative_2013,limmer_fluctuations_2013,moore_growing_2009,montero_de_hijes_viscosity_2018}. Those features are found in simulations where a liquid-liquid transition occurs in deeply metastable water~\cite{becker_fractional_2006,kumar_relation_2007,montero_de_hijes_viscosity_2018,mazza_connection_2007}, as well as in simulations without such a transition~\cite{moore_growing_2009}. Phenomenological two-state models~\cite{anisimov_thermodynamics_2018}, which treats water as a non-ideal mixture of inter-converting species, whose fraction depends on temperature and pressure, successfully reproduce thermodynamic and dynamic data of real~\cite{holten_equation_2014,singh_pressure_2017,caupin_thermodynamics_2019,duska_water_2020} and simulated~\cite{biddle_two-structure_2017,montero_de_hijes_viscosity_2018} supercooled water. They are compatible with the existence of a liquid-liquid transition, but this is not necessary~\cite{caupin_minimal_2021}. The vicinity of a critical point terminating a liquid-liquid transition, or the molecular fluctuations between two different local structures, could induce dynamic heterogeneities in bulk light and heavy water, and provide a rationale to explain our observations.

\section*{Acknowledgments}
We acknowledge support from Agence Nationale de la Recherche, grant number ANR-19-CE30-0035-01.

\appendix*

\section{Smoothed values of the viscosity}

\begingroup
\squeezetable
\begin{table*}[h!]
\caption{\label{Ragueneau_Smoothed} Smoothed values of the viscosity of H$_2$O (resp. D$_2$O) as functions of the temperature: $\eta=\eta_0(T-T_0)^{-\gamma}$ with $\eta_0=137.46\,\mathrm{\mu Pa\,s}$, $T_s=225.9151\,\mathrm{K}$ and $\gamma=1.6383$ (resp. $\eta_0=132.12\,\mathrm{\mu Pa\,s}$, $T_s=230.9681\,\mathrm{K}$ and $\gamma=1.7061$).}
\begin{tabular}{|c|cc|cc|}
\hline
					 &		\multicolumn{2}{c|}{H$_2$O}											&		\multicolumn{2}{c|}{D$_2$O}	\\
\hline
T ($\mathrm{K}$) 	   & 	 $\eta$ ($\mathrm{mPa\,s}$) 	 & 	 Relative uncertainty ($\%$) 	  & 	 $\eta$ ($\mathrm{mPa\,s}$) 	 & 	 Relative uncertainty ($\%$) 	 \\ 
\hline
 240.15  	 & 	 12.74  	 				 & 	 2.1 	 													&											 &																 \\ 
 241.15  	 & 	 11.40  	 				 & 	 2.1 	 													&											 &																 \\ 
 242.15  	 & 	 10.27  					 & 	 2.0 	 													&											 &																 \\ 
 243.15  	 & 	 9.31  						 & 	 1.9 	 													&											 &																 \\ 
 243.70  	 & 										 &	 																&			18.55					 	 & 	 7.0 													 \\
 244.15  	 & 	 8.49  						 & 	 1.9 	 													& 	  17.49  	 			   & 	 7.0 	 												 \\
 245.15  	 & 	 7.78  						 & 	 1.8 	 													& 	  15.44  			   	 & 	 7.0 	 \\ 
 246.15  	 & 	 7.16  						 & 	 1.8 	 													&     13.74  	    	   & 	 7.0 	 \\ 
 247.15  	 & 	 6.62  	 	  			 & 	 1.8 	 													& 	  12.32  			   	 & 	 7.0 	 \\ 
 248.15  	 & 	 6.14  						 & 	 1.7 	 													& 	  11.13  			   	 & 	 7.0 	 \\ 
 249.15  	 & 	 5.71  						 & 	 1.7 	 													& 	  10.10  			   	 & 	 7.0 	 \\  
 250.15  	 & 	 5.33  						 & 	 1.7 	 													& 	  9.22  			   	 & 	 4.0 	 \\
 251.15  	 & 	 4.99  	 					 & 	 1.7 	 													& 	  8.45  			   	 & 	 4.0 	 \\  
 252.15  	 & 	 4.68  	 					 & 	 1.7 	 													& 	  7.78  			   	 & 	 4.0 	 \\ 
 253.15  	 & 	 4.40  	 					 & 	 1.7 	 													& 	  7.20 			    	 & 	 4.0 	 \\
 254.15  	 & 	 4.15  	 					 & 	 1.7 	 													& 	  6.67  	     	   & 	 4.0 	 \\ 
 255.15  	 & 	 3.92  	 					 & 	 1.6 	 													& 	  6.21  	     	   & 	 3.0 	 \\ 
 256.15  	 & 	 3.71  	 					 & 	 1.6 	 													& 	  5.80  	     	   & 	 3.0 	 \\
 257.15  	 & 	 3.52  	 					 & 	 1.6 	 													& 	  5.42  	     	   & 	 3.0 	 \\
 258.15  	 & 	 3.34  	 					 & 	 1.6 	 													& 	  5.09  	     	   & 	 3.0 	 \\
 259.15  	 & 	 3.18  						 & 	 1.6 	 													& 	  4.78  	     	   & 	 3.0 	 \\
 260.15  	 & 	 3.03  						 & 	 1.6 	 													& 	  4.51  	     	   & 	 2.0 	 \\
 261.15  	 & 	 2.89 						 & 	 1.6 	 													& 	  4.25  	     	   & 	 2.0 	 \\
 262.15  	 & 	 2.76 					 	 & 	 1.6 	 													& 	  4.02  	     	   & 	 2.0 	 \\
 263.15  	 & 	 2.64  						 & 	 1.6 	 													& 	  3.81  	     	   & 	 2.0 	 \\
 264.15  	 & 	 2.52  						 & 	 1.6 	 													& 	  3.62  	     	   & 	 2.0 	 \\
 265.15  	 & 	 2.42  						 & 	 1.6 	 													& 	  3.44  	     	   & 	 2.0 	 \\
 266.15  	 & 	 2.32  						 & 	 1.6 	 													& 	  3.28  	     	   & 	 2.0 	 \\
 267.15  	 & 	 2.23  						 & 	 1.6 	 													& 	  3.12  	     	   & 	 2.0 	 \\ 
 268.15  	 & 	 2.14  						 & 	 1.6 	 													& 	  2.98  	     	   & 	 2.0 	 \\ 
 269.15  	 & 	 2.06  						 & 	 1.6 	 													& 	  2.85  	     	   & 	 2.0 	 \\
 270.15  	 & 	 1.99  						 & 	 1.6 	 													& 	  2.73  	    	   & 	 2.0 	 \\
 271.15  	 & 	 1.92  						 & 	 1.6 	 													& 	  2.61      	   	 & 	 2.0 	 \\
 272.15  	 & 	 1.85 					 	 & 	 1.6 	 													& 	  2.50      	   	 & 	 2.0 	 \\
 273.15  	 & 	 1.79  						 & 	 1.6 	 													& 	  2.40      	   	 & 	 2.0 	 \\
 274.15  	 & 	 1.73  						 & 	 1.6 	 													& 	  2.31  	    	   & 	 2.0 	 \\
 275.15  	 & 	 1.67  						 & 	 1.6 	 													& 	  2.22      	   	 & 	 2.0 	 \\ 
 276.15  	 & 	 1.61  						 & 	 1.6 	 													& 	  2.14      	   	 & 	 2.0 	 \\
 277.15  	 & 	 1.56  					   & 	 1.6 	 													& 	  2.06      	   	 & 	 2.0 	 \\
 278.15  	 & 	 1.51  						 & 	 1.6 	 													& 	  1.99      	     & 	 2.0 	 \\
 279.15  	 & 	 1.47  						 & 	 1.6 	 													& 	  1.92  	    	   & 	 2.0 	 \\
 280.15  	 & 	 1.42  						 & 	 1.6 	 													& 	  1.85  	    	   & 	 2.0 	 \\ 
 281.15  	 & 	 1.38  						 & 	 1.6 	 													& 	  1.79      	   	 & 	 2.0 	 \\
 282.15  	 & 	 1.34  						 & 	 1.6 	 													& 	  1.73      	   	 & 	 2.0 	 \\
 283.15  	 & 	 1.30 						 & 	 1.6 	 													& 	  1.67      	   	 & 	 2.0 	 \\
 284.15  	 & 	 1.27  						 & 	 1.6 	 													& 	  1.62      	   	 & 	 2.0 	 \\
 285.15  	 & 	 1.23  					   & 	 1.6 	 													& 	  1.57      	   	 & 	 2.0 	 \\
 286.15  	 & 	 1.20  						 & 	 1.6 	 													& 	  1.52      	   	 & 	 2.0 	 \\
 287.15  	 & 	 1.17  						 & 	 1.6 	 													& 	  1.47      	   	 & 	 2.0 	 \\
 288.15  	 & 	 1.14  						 & 	 1.6 	 													& 	  1.43      	   	 & 	 2.0 	 \\
 289.15  	 & 	 1.11  						 & 	 1.6 	 													& 	  1.39      	   	 & 	 2.0 	 \\ 
 290.15  	 & 	 1.08  						 & 	 1.5 	 													& 	  1.35      	   	 & 	 2.0 	 \\
 291.15  	 & 	 1.05  						 & 	 1.5 	 													& 	  1.31      	   	 & 	 2.0 	 \\
 292.15  	 & 	 1.03  						 & 	 1.5 	 													& 	  1.27      	   	 & 	 2.0 	 \\
 293.15  	 & 	 1.00  						 & 	 1.5 	 													& 	  1.24      	   	 & 	 2.0 	 \\
\hline
\end{tabular}
\end{table*}
\endgroup

%

\end{document}



\title{Shear viscosity and Stokes-Einstein violation in supercooled light and heavy water \\ Supplemental Material}


\author{P. Ragueneau, F. Caupin and B. Issenmann}
\email[]{frederic.caupin@univ-lyon1.fr, bruno.issenmann@univ-lyon1.fr	}
\affiliation{Institut Lumi\`ere Mati\`ere, Universit\'e de Lyon, Universit\'e Claude Bernard Lyon 1, CNRS, F-69622, Villeurbanne, France}

\date{\today}

\maketitle

\section{Reorientation of water molecules and Stokes-Einstein-Debye relation}

The Debye model~\cite{debye_polar_1929} which leads to the Stokes-Einstein-Debye (SED) relation assumes rotational Brownian motion of the reorienting object with a diffusion coefficient $D_\mathrm{rot}$. This holds if the orientation changes as in a random walk, by small angular steps. With $\vec{u}$ a unit vector describing the orientation, one defines the $l$-th order correlation function:
\begin{equation}
C_l (t) = \langle P_l[\vec{u}(0)\cdot\vec{u}(t)]\rangle \, ,
\end{equation}
where $P_l$ is the $l$-th order Legendre polynomial and the angle brackets denote an ensemble average. The Debye theory predicts that the correlation functions decay exponentially~\cite{debye_polar_1929}:
\begin{equation}
C_l (t) = \exp (- t/\tau_l) \quad \mathrm{with}\quad \tau_l = \frac{1}{l(l+1)D_\mathrm{rot}} \, .
\end{equation}
The generalized Stokes-Einstein-Debye relation would thus be:
\begin{equation}
\frac{\eta}{l (l+1) \, T \tau_l} = \mathrm{cst}\,,
\label{eq:SEDl}
\end{equation}
with the same constant for all values of $l$.

NMR gives access to the integral rotational correlation time~\cite{qvist_rotational_2012}:
\begin{equation}
\tau_\theta = \int_0^{+\infty} C_2(t) \,\mathrm{d}t \, ,
\end{equation}
which is the quantity we consider in the present work. In the case of a Debye process, the rotational diffusion coefficient would thus be $D_\mathrm{rot} = 1/(6\,\tau_\theta )$. Unfortunately, this simple picture does not apply to reorientation in water. Simulations have shown that a water molecule reorients by large-amplitude angular jumps~\cite{laage_molecular_2006}, although the jump angle and the detailed description of the process vary among authors~\cite{laage_molecular_2006,qvist_rotational_2012}. The rotating water molecule breaks a hydrogen bond with an over-coordinated neighbor in its first shell, to form another bond with an under-coordinated neighbor in its second shell. Therefore, the NMR data should not be interpreted in terms of $D_\mathrm{rot}$.

Recent simulations with the TIP4P/2005 water model have investigated how Eq.~\ref{eq:SEDl} is violated~\cite{kawasaki_spurious_2019}. It was found that $\eta/(T\tau_l )$ increases when the temperature decreases, and behaves differently for different $l$. The dependance on the degree of Legendre polynomials led the authors to conclude that the SED violation is spurious. Nevertheless, it is remarkable that, for a given $l$, these simulations find $\eta/(T\tau_l )$ to be nearly temperature independent above $250\,\mathrm{K}$. This is consistent with what is experimentally observed for $\eta/(T\tau_\theta )$. Therefore, while the Debye model is not applicable to the processes governing molecular reorientation in water, the SED relation remains a useful equation to assess the scaling between the experimentally accessible quantities, $\eta$ and $\tau_\theta$.

\newpage
\begin{widetext}
\section{Tables}

\begin{longtable*}{|c|c|c|c|c|c|c|c|c|c|c|c|}
\caption{Raw values of the viscosity of H$_2$O}
\\
\hline
run number 	 & 	 T (K) 	 & 	 $\eta\, (\mathrm{mPa\,s})$ 	 & 	 $\eta\, (\mathrm{mPa\,s})$ 	 & 	 $\eta\, (\mathrm{mPa\,s})$ 	 & 	 $\eta\, (\mathrm{mPa\,s})$ 	 & 	 $\eta\, (\mathrm{mPa\,s})$ 	 & 	 $\eta\, (\mathrm{mPa\,s})$ 	 & 	 $\eta\, (\mathrm{mPa\,s})$ 	 & 	 $\eta\, (\mathrm{mPa\,s})$ 	 & 	 $\eta\, (\mathrm{mPa\,s})$ 	 & 	 $\eta\, (\mathrm{mPa\,s})$   \\ 	  
\hline
\endhead 
\hline
\endfoot
\multirow{2}{*}{1}
  	 & 	 293.15  	 & 	 1.01  	 & 	 1.00 	 & 	 1.00  	 & 	 1.00  	 & 	 0.99  	 & 	 1.01  	 & 	      	 & 	      	 & 	      	 & 	      	 \\
  	 & 	 250.51  	 & 	 5.12  	 & 	 5.21 	 & 	 5.17  	 & 	      	 & 	      	 & 	      	 & 	      	 & 	      	 & 	      	 & 	      	 \\
\hline
\multirow{2}{*}{2}
  	 & 	 293.15  	 & 	 1.01  	 & 	 1.00 	 & 	 1.00  	 & 	 1.01  	 & 	 1.00  	 & 	 1.02  	 & 	 1.01  	 & 	 0.99  	 & 	 0.98  	 & 	      	 \\ 
  	 & 	 250.51  	 & 	 5.28  	 & 	 5.24 	 & 	      	 & 	      	 & 	      	 & 	      	 & 	      	 & 	      	 & 	      	 & 	      	 \\ 
\hline
\multirow{3}{*}{3}
  	 & 	 293.15  	 & 	 1.03  	 & 	 1.00 	 & 	 1.00  	 & 	 0.97  	 & 	 1.02  	 & 	 1.01  	 & 	 1.00  	 & 	 1.00  	 & 	 1.00  	 & 	 0.99  	 \\ 
  	 & 	 254.41  	 & 	 4.10  	 & 	 4.10 	 & 	 4.15  	 & 	      	 & 	      	 & 	      	 & 	      	 & 	      	 & 	      	 & 	      	 \\ 
  	 & 	 250.57  	 & 	 5.31  	 & 	 5.24 	 & 	 5.16  	 & 	      	 & 	      	 & 	      	 & 	      	 & 	      	 & 	      	 & 	      	 \\ 
\hline
\multirow{4}{*}{4}
  	 & 	 293.15  	 & 	 1.01  	 & 	 0.99 	 & 	 1.00  	 & 	 1.00  	 & 	 1.01  	 & 	 1.01  	 & 	 1.00  	 & 	 1.01  	 & 	 0.99  	 & 	 0.99  	 \\ 
  	 & 	 254.36  	 & 	 4.13  	 & 	 4.05 	 & 	 4.12  	 & 	      	 & 	      	 & 	      	 & 	      	 & 	      	 & 	      	 & 	      	 \\ 
  	 & 	 252.43  	 & 	 4.38  	 & 	 4.41 	 & 	 4.63  	 & 	      	 & 	      	 & 	      	 & 	      	 & 	      	 & 	      	 & 	      	 \\ 
  	 & 	 250.51  	 & 	 5.19  	 & 	 4.91 	 & 	 5.17  	 & 	      	 & 	      	 & 	      	 & 	      	 & 	      	 & 	      	 & 	      	 \\ 
\hline
\multirow{13}{*}{5}
  	 & 	 293.15  	 & 	 0.99  	 & 	 1.00 	 & 	 1.02  	 & 	 1.00  	 & 	 1.01  	 & 	 0.99  	 & 	 0.99  	 & 	 1.00  	 & 	 1.00  	 & 	 1.01  	 \\ 
  	 & 	 254.56  	 & 	 3.98  	 & 	 3.92 	 & 	 3.96  	 & 	      	 & 	      	 & 	      	 & 	      	 & 	      	 & 	      	 & 	      	 \\ 
  	 & 	 252.66  	 & 	 4.62  	 & 	 4.52 	 & 	 4.60  	 & 	      	 & 	      	 & 	      	 & 	      	 & 	      	 & 	      	 & 	      	 \\ 
  	 & 	 250.75  	 & 	 5.12  	 & 	 5.12 	 & 	 5.19  	 & 	      	 & 	      	 & 	      	 & 	      	 & 	      	 & 	      	 & 	      	 \\ 
  	 & 	 248.85  	 & 	 5.70  	 & 	 5.86 	 & 	 5.76  	 & 	      	 & 	      	 & 	      	 & 	      	 & 	      	 & 	      	 & 	      	 \\ 
  	 & 	 246.94  	 & 	 6.62  	 & 	 6.56 	 & 	 6.56  	 & 	      	 & 	      	 & 	      	 & 	      	 & 	      	 & 	      	 & 	      	 \\ 
  	 & 	 245.03  	 & 	 7.94  	 & 	 8.11 	 & 	 7.60  	 & 	      	 & 	      	 & 	      	 & 	      	 & 	      	 & 	      	 & 	      	 \\ 
  	 & 	 244.08  	 & 	 8.53  	 & 	 8.34 	 & 	 8.68  	 & 	      	 & 	      	 & 	      	 & 	      	 & 	      	 & 	      	 & 	      	 \\ 
  	 & 	 243.13  	 & 	 9.61  	 & 	 9.26 	 & 	 9.22  	 & 	      	 & 	      	 & 	      	 & 	      	 & 	      	 & 	      	 & 	      	 \\ 
  	 & 	 242.17  	 & 	 10.17  	 & 	 10.24 	 & 	 10.24  	 & 	      	 & 	      	 & 	      	 & 	      	 & 	      	 & 	      	 & 	      	 \\ 
  	 & 	 241.22  	 & 	 11.44  	 & 	 11.35 	 & 	 11.22  	 & 	      	 & 	      	 & 	      	 & 	      	 & 	      	 & 	      	 & 	      	 \\ 
  	 & 	 240.74  	 & 	 11.42  	 & 	 11.80 	 & 	 11.91  	 & 	      	 & 	      	 & 	      	 & 	      	 & 	      	 & 	      	 & 	      	 \\ 
  	 & 	 240.27  	 & 	 12.97  	 & 	 12.34 	 & 	      	 & 	      	 & 	      	 & 	      	 & 	      	 & 	      	 & 	      	 & 	      	 \\ 
\hline
\multirow{7}{*}{6}
  	 & 	 293.15  	 & 	 0.99  	 & 	 1.01 	 & 	 1.00  	 & 	 1.00  	 & 	 1.01  	 & 	 1.00  	 & 	 1.00  	 & 	 0.99  	 & 	 1.00  	 & 	 1.00  	 \\ 
  	 & 	 288.10  	 & 	 1.14  	 & 	 1.13 	 & 	 1.13  	 & 	      	 & 	      	 & 	      	 & 	      	 & 	      	 & 	      	 & 	      	 \\ 
  	 & 	 283.31  	 & 	 1.31  	 & 	 1.29 	 & 	 1.33  	 & 	      	 & 	      	 & 	      	 & 	      	 & 	      	 & 	      	 & 	      	 \\ 
  	 & 	 278.52  	 & 	 1.51  	 & 	 1.50 	 & 	 1.51  	 & 	      	 & 	      	 & 	      	 & 	      	 & 	      	 & 	      	 & 	      	 \\ 
  	 & 	 273.72  	 & 	 1.75  	 & 	 1.78 	 & 	 1.78  	 & 	      	 & 	      	 & 	      	 & 	      	 & 	      	 & 	      	 & 	      	 \\ 
  	 & 	 268.93  	 & 	 2.09  	 & 	 2.11 	 & 	 2.10  	 & 	      	 & 	      	 & 	      	 & 	      	 & 	      	 & 	      	 & 	      	 \\ 
  	 & 	 264.14  	 & 	 2.54  	 & 	 2.59 	 & 	 2.62  	 & 	      	 & 	      	 & 	      	 & 	      	 & 	      	 & 	      	 & 	      	 \\ 
\pagebreak
\multirow{11}{*}{6}
  	 & 	 259.35  	 & 	 3.21  	 & 	 3.21 	 & 	 3.24  	 & 	      	 & 	      	 & 	      	 & 	      	 & 	      	 & 	      	 & 	      	 \\ 
  	 & 	 254.56  	 & 	 4.11  	 & 	 4.08 	 & 	 4.09  	 & 	      	 & 	      	 & 	      	 & 	      	 & 	      	 & 	      	 & 	      	 \\ 
  	 & 	 252.64  	 & 	 4.58  	 & 	 4.68 	 & 	 4.56  	 & 	      	 & 	      	 & 	      	 & 	      	 & 	      	 & 	      	 & 	      	 \\ 
  	 & 	 250.73  	 & 	 5.21  	 & 	 5.25 	 & 	 5.20  	 & 	      	 & 	      	 & 	      	 & 	      	 & 	      	 & 	      	 & 	      	 \\ 
  	 & 	 248.81  	 & 	 5.85  	 & 	 5.94 	 & 	 5.82  	 & 	      	 & 	      	 & 	      	 & 	      	 & 	      	 & 	      	 & 	      	 \\ 
  	 & 	 246.89  	 & 	 6.87  	 & 	 6.86 	 & 	 6.87  	 & 	      	 & 	      	 & 	      	 & 	      	 & 	      	 & 	      	 & 	      	 \\ 
  	 & 	 244.98  	 & 	 7.83  	 & 	 7.94 	 & 	 7.90  	 & 	      	 & 	      	 & 	      	 & 	      	 & 	      	 & 	      	 & 	      	 \\ 
  	 & 	 244.02  	 & 	 8.62  	 & 	 8.62 	 & 	 8.77  	 & 	      	 & 	      	 & 	      	 & 	      	 & 	      	 & 	      	 & 	      	 \\ 
  	 & 	 243.06  	 & 	 9.49  	 & 	 9.47 	 & 	 9.40  	 & 	      	 & 	      	 & 	      	 & 	      	 & 	      	 & 	      	 & 	      	 \\ 
  	 & 	 242.10  	 & 	 10.36  	 & 	 10.34 	 & 	 10.34  	 & 	      	 & 	      	 & 	      	 & 	      	 & 	      	 & 	      	 & 	      	 \\ 
  	 & 	 241.14  	 & 	 11.51  	 & 	 11.51 	 & 	 11.40  	 & 	      	 & 	      	 & 	      	 & 	      	 & 	      	 & 	      	 & 	      	 \\ 
\hline
\multirow{27}{*}{7}
  	 & 	 293.15  	 & 	 0.99  	 & 	 1.00 	 & 	 1.01  	 & 	 1.00  	 & 	 1.02  	 & 	 1.00  	 & 	 1.02  	 & 	 0.99  	 & 	 1.00  	 & 	 0.99  	 \\ 
  	 & 	 287.88  	 & 	 1.18  	 & 	     	 & 	      	 & 	      	 & 	      	 & 	      	 & 	      	 & 	      	 & 	      	 & 	      	 \\ 
  	 & 	 283.13  	 & 	 1.32  	 & 	 1.26 	 & 	      	 & 	      	 & 	      	 & 	      	 & 	      	 & 	      	 & 	      	 & 	      	 \\ 
  	 & 	 278.38  	 & 	 1.46  	 & 	 1.45 	 & 	 1.45  	 & 	      	 & 	      	 & 	      	 & 	      	 & 	      	 & 	      	 & 	      	 \\ 
  	 & 	 273.63  	 & 	 1.72  	 & 	 1.71 	 & 	 1.71  	 & 	      	 & 	      	 & 	      	 & 	      	 & 	      	 & 	      	 & 	      	 \\ 
  	 & 	 268.87  	 & 	 2.06  	 & 	 2.03 	 & 	 2.01  	 & 	      	 & 	      	 & 	      	 & 	      	 & 	      	 & 	      	 & 	      	 \\ 
  	 & 	 264.12  	 & 	 2.52  	 & 	 2.49 	 & 	 2.51  	 & 	      	 & 	      	 & 	      	 & 	      	 & 	      	 & 	      	 & 	      	 \\ 
  	 & 	 261.74  	 & 	 2.81  	 & 	 2.75 	 & 	 2.79  	 & 	      	 & 	      	 & 	      	 & 	      	 & 	      	 & 	      	 & 	      	 \\ 
  	 & 	 259.37  	 & 	 3.08  	 & 	 3.10 	 & 	 3.09  	 & 	      	 & 	      	 & 	      	 & 	      	 & 	      	 & 	      	 & 	      	 \\ 
  	 & 	 256.99  	 & 	 3.60  	 & 	 3.53 	 & 	 3.51  	 & 	      	 & 	      	 & 	      	 & 	      	 & 	      	 & 	      	 & 	      	 \\ 
  	 & 	 254.61  	 & 	 4.07  	 & 	 4.06 	 & 	 4.03  	 & 	      	 & 	      	 & 	      	 & 	      	 & 	      	 & 	      	 & 	      	 \\ 
  	 & 	 253.66  	 & 	 4.26  	 & 	 4.26 	 & 	 4.29  	 & 	      	 & 	      	 & 	      	 & 	      	 & 	      	 & 	      	 & 	      	 \\ 
  	 & 	 252.71  	 & 	 4.50  	 & 	 4.47 	 & 	 4.52  	 & 	      	 & 	      	 & 	      	 & 	      	 & 	      	 & 	      	 & 	      	 \\ 
  	 & 	 251.76  	 & 	 4.83  	 & 	 4.71 	 & 	 4.72  	 & 	      	 & 	      	 & 	      	 & 	      	 & 	      	 & 	      	 & 	      	 \\ 
  	 & 	 250.81  	 & 	 5.06  	 & 	 5.13 	 & 	 5.05  	 & 	      	 & 	      	 & 	      	 & 	      	 & 	      	 & 	      	 & 	      	 \\ 
  	 & 	 249.86  	 & 	 5.47  	 & 	 5.42 	 & 	 5.39  	 & 	      	 & 	      	 & 	      	 & 	      	 & 	      	 & 	      	 & 	      	 \\ 
  	 & 	 248.91  	 & 	 5.75  	 & 	 5.78 	 & 	 5.81  	 & 	      	 & 	      	 & 	      	 & 	      	 & 	      	 & 	      	 & 	      	 \\ 
  	 & 	 247.96  	 & 	 6.27  	 & 	 6.30 	 & 	 6.18  	 & 	      	 & 	      	 & 	      	 & 	      	 & 	      	 & 	      	 & 	      	 \\ 
  	 & 	 247.01  	 & 	 6.64  	 & 	 6.64 	 & 	 6.68  	 & 	      	 & 	      	 & 	      	 & 	      	 & 	      	 & 	      	 & 	      	 \\ 
  	 & 	 246.06  	 & 	 7.31  	 & 	 7.15 	 & 	 7.22  	 & 	      	 & 	      	 & 	      	 & 	      	 & 	      	 & 	      	 & 	      	 \\ 
  	 & 	 245.11  	 & 	 7.78  	 & 	 7.86 	 & 	 7.90  	 & 	      	 & 	      	 & 	      	 & 	      	 & 	      	 & 	      	 & 	      	 \\ 
  	 & 	 244.16  	 & 	 8.43  	 & 	 8.60 	 & 	 8.47  	 & 	      	 & 	      	 & 	      	 & 	      	 & 	      	 & 	      	 & 	      	 \\ 
  	 & 	 243.68  	 & 	 8.84  	 & 	 8.89 	 & 	 8.81  	 & 	      	 & 	      	 & 	      	 & 	      	 & 	      	 & 	      	 & 	      	 \\ 
  	 & 	 243.21  	 & 	 9.16  	 & 	 9.13 	 & 	 9.32  	 & 	      	 & 	      	 & 	      	 & 	      	 & 	      	 & 	      	 & 	      	 \\ 
  	 & 	 242.73  	 & 	 9.55  	 & 	 9.85 	 & 	 9.71  	 & 	      	 & 	      	 & 	      	 & 	      	 & 	      	 & 	      	 & 	      	 \\ 
  	 & 	 242.26  	 & 	 10.34  	 & 	 10.11 	 & 	 10.24  	 & 	      	 & 	      	 & 	      	 & 	      	 & 	      	 & 	      	 & 	      	 \\ 
  	 & 	 241.78  	 & 	 10.61  	 & 	     	 & 	      	 & 	      	 & 	      	 & 	      	 & 	      	 & 	      	 & 	      	 & 	      	 \\ 
\hline
\multirow{8}{*}{8}
  	 & 	 293.15  	 & 	 0.99  	 & 	 1.00 	 & 	 1.01  	 & 	 1.00  	 & 	 1.01  	 & 	 0.99  	 & 	 1.01  	 & 	 0.99  	 & 	 1.02  	 & 	 1.00  	 \\ 
  	 & 	 283.36  	 & 	 1.31  	 & 	 1.31 	 & 	 1.30  	 & 	      	 & 	      	 & 	      	 & 	      	 & 	      	 & 	      	 & 	      	 \\ 
  	 & 	 273.73  	 & 	 1.79  	 & 	 1.76 	 & 	 1.78  	 & 	      	 & 	      	 & 	      	 & 	      	 & 	      	 & 	      	 & 	      	 \\ 
  	 & 	 264.09  	 & 	 2.53  	 & 	 2.56 	 & 	 2.55  	 & 	      	 & 	      	 & 	      	 & 	      	 & 	      	 & 	      	 & 	      	 \\ 
  	 & 	 254.46  	 & 	 4.20  	 & 	 4.10 	 & 	 4.04  	 & 	      	 & 	      	 & 	      	 & 	      	 & 	      	 & 	      	 & 	      	 \\ 
  	 & 	 249.64  	 & 	 5.57  	 & 	 5.57 	 & 	      	 & 	      	 & 	      	 & 	      	 & 	      	 & 	      	 & 	      	 & 	      	 \\ 
  	 & 	 244.82  	 & 	 7.94  	 & 	 7.87 	 & 	      	 & 	      	 & 	      	 & 	      	 & 	      	 & 	      	 & 	      	 & 	      	 \\ 
  	 & 	 242.89  	 & 	 9.51  	 & 	     	 & 	      	 & 	      	 & 	      	 & 	      	 & 	      	 & 	      	 & 	      	 & 	      	 \\ 
\hline
\multirow{9}{*}{9}
  	 & 	 293.15  	 & 	 1.00  	 & 	 1.00 	 & 	 1.00  	 & 	 1.00  	 & 	 1.01  	 & 	 0.99  	 & 	 0.99  	 & 	 1.01  	 & 	 1.01  	 & 	 1.00  	 \\ 
  	 & 	 283.28  	 & 	 1.30  	 & 	 1.31 	 & 	 1.31  	 & 	      	 & 	      	 & 	      	 & 	      	 & 	      	 & 	      	 & 	      	 \\ 
  	 & 	 273.72  	 & 	 1.75  	 & 	 1.75 	 & 	 1.73  	 & 	      	 & 	      	 & 	      	 & 	      	 & 	      	 & 	      	 & 	      	 \\ 
  	 & 	 264.17  	 & 	 2.56  	 & 	 2.58 	 & 	 2.54  	 & 	      	 & 	      	 & 	      	 & 	      	 & 	      	 & 	      	 & 	      	 \\ 
  	 & 	 254.61  	 & 	 4.09  	 & 	 4.09 	 & 	 3.97  	 & 	      	 & 	      	 & 	      	 & 	      	 & 	      	 & 	      	 & 	      	 \\ 
  	 & 	 249.83  	 & 	 5.56  	 & 	 5.54 	 & 	 5.44  	 & 	      	 & 	      	 & 	      	 & 	      	 & 	      	 & 	      	 & 	      	 \\ 
  	 & 	 245.05  	 & 	 7.72  	 & 	 7.92 	 & 	      	 & 	      	 & 	      	 & 	      	 & 	      	 & 	      	 & 	      	 & 	      	 \\ 
  	 & 	 243.14  	 & 	 9.44  	 & 	 9.37 	 & 	 9.20  	 & 	      	 & 	      	 & 	      	 & 	      	 & 	      	 & 	      	 & 	      	 \\ 
  	 & 	 241.23  	 & 	 11.45  	 & 	 11.21 	 & 	      	 & 	      	 & 	      	 & 	      	 & 	      	 & 	      	 & 	      	 & 	      	 \\ 
\hline
\end{longtable*}

\newpage

\begin{longtable*}{|c|c|c|c|c|c|c|c|c|c|c|c|c|}
\caption{Raw values of the viscosity of H$_2$O$-$D$_2$O mixtures}
\\
\hline
run number 	 & 	  $x_\mathrm{D_{2}O}$ ($\%$) 	 & 	 T (K) 	 & 	 $\eta\, (\mathrm{mPa\,s})$ 	 & 	 $\eta\, (\mathrm{mPa\,s})$ 	 & 	 $\eta\, (\mathrm{mPa\,s})$ 	 & 	 $\eta\, (\mathrm{mPa\,s})$ 	 & 	 $\eta\, (\mathrm{mPa\,s})$ 	 & 	 $\eta\, (\mathrm{mPa\,s})$ 	 & 	 $\eta\, (\mathrm{mPa\,s})$ 	 & 	 $\eta\, (\mathrm{mPa\,s})$ 	 & 	 $\eta\, (\mathrm{mPa\,s})$ 	 & 	 $\eta\, (\mathrm{mPa\,s})$   	 \\
\hline
\endhead 
\hline
\endfoot
\multirow{10}{*}{1}	&	\multirow{10}{*}{93.39}	
 	 & 	 293.15  	 & 	 1.25  	 & 	 1.23 	 & 	 1.24  	 & 	 1.22  	 & 	 1.24  	 & 	 1.22  	 & 	 1.22  	 & 	 1.23  	 & 	 1.23  	 & 	 1.24  	 \\ 
 	 &&	 278.47  	 & 	 1.92  	 & 	 1.92 	 & 	 1.91  	 & 	      	 & 	      	 & 	      	 & 	      	 & 	      	 & 	      	 & 	      	 \\ 
 	&& 	 273.63  	 & 	 2.30  	 & 	 2.31 	 & 	 2.29  	 & 	      	 & 	      	 & 	      	 & 	      	 & 	      	 & 	      	 & 	      	 \\ 
 	&& 	 268.80  	 & 	 2.81  	 & 	 2.81 	 & 	 2.83  	 & 	      	 & 	      	 & 	      	 & 	      	 & 	      	 & 	      	 & 	      	 \\ 
 	&& 	 263.97  	 & 	 3.61  	 & 	 3.59 	 & 	 3.58  	 & 	      	 & 	      	 & 	      	 & 	      	 & 	      	 & 	      	 & 	      	 \\ 
 	&& 	 259.14  	 & 	 4.71  	 & 	 4.71 	 & 	 4.72  	 & 	      	 & 	      	 & 	      	 & 	      	 & 	      	 & 	      	 & 	      	 \\ 
 	&& 	 256.24  	 & 	 5.67  	 & 	 5.66 	 & 	 5.62  	 & 	      	 & 	      	 & 	      	 & 	      	 & 	      	 & 	      	 & 	      	 \\ 
 	&& 	 254.30  	 & 	 6.45  	 & 	 6.47 	 & 	 6.53  	 & 	      	 & 	      	 & 	      	 & 	      	 & 	      	 & 	      	 & 	      	 \\ 
 	&& 	 252.37  	 & 	 7.34  	 & 	 7.48 	 & 	 7.58  	 & 	      	 & 	      	 & 	      	 & 	      	 & 	      	 & 	      	 & 	      	 \\ 
 	&& 	 251.40  	 & 	 8.13  	 & 	 8.05 	 & 	 8.11  	 & 	      	 & 	      	 & 	      	 & 	      	 & 	      	 & 	      	 & 	      	 \\ 
 	&& 	 250.44  	 & 	 8.71  	 & 	 8.87 	 & 	 8.75  	 & 	      	 & 	      	 & 	      	 & 	      	 & 	      	 & 	      	 & 	      	 \\ 
	&& 	 249.47  	 & 	 9.53  	 & 	 9.53 	 & 	 9.69  	 & 	      	 & 	      	 & 	      	 & 	      	 & 	      	 & 	      	 & 	      	 \\ 
 	&& 	 248.50  	 & 	 10.74  	 & 	 10.63 	 & 	 10.49  	 & 	      	 & 	      	 & 	      	 & 	      	 & 	      	 & 	      	 & 	      	 \\ 
 	&& 	 247.54  	 & 	 11.37  	 & 	 11.23 	 & 	 11.69  	 & 	      	 & 	      	 & 	      	 & 	      	 & 	      	 & 	      	 & 	      	 \\ 
 	&& 	 246.57  	 & 	 12.92  	 & 	 13.14 	 & 	 12.88  	 & 	      	 & 	      	 & 	      	 & 	      	 & 	      	 & 	      	 & 	      	 \\ 
 	&& 	 246.09  	 & 	 13.62  	 & 	 13.51 	 & 	 13.74  	 & 	      	 & 	      	 & 	      	 & 	      	 & 	      	 & 	      	 & 	      	 \\ 
	&& 	 245.61  	 & 	 14.48  	 & 	 14.35 	 & 	 14.62  	 & 	      	 & 	      	 & 	      	 & 	      	 & 	      	 & 	      	 & 	      	 \\ 
 	&& 	 245.12  	 & 	 15.20  	 & 	 15.12 	 & 	 15.48  	 & 	      	 & 	      	 & 	      	 & 	      	 & 	      	 & 	      	 & 	      	 \\ 
 	&& 	 244.64  	 & 	 16.15  	 & 	 16.41 	 & 	 16.39  	 & 	      	 & 	      	 & 	      	 & 	      	 & 	      	 & 	      	 & 	      	 \\ 
	&& 	 244.16  	 & 	 17.34  	 & 	 16.94 	 & 	 17.25  	 & 	      	 & 	      	 & 	      	 & 	      	 & 	      	 & 	      	 & 	      	 \\ 
 	&& 	 243.67  	 & 	 18.07  	 & 	     	 & 	      	 & 	      	 & 	      	 & 	      	 & 	      	 & 	      	 & 	      	 & 	      	 \\ 
\hline 
\multirow{5}{*}{2}	&	\multirow{5}{*}{92.82}	
 & 	 293.15  	 & 	 1.21  	 & 	 1.24 	 & 	 1.23  	 & 	 1.24  	 & 	 1.23  	 & 	 1.23  	 & 	 1.25  	 & 	 1.23  	 & 	 1.22  	 & 	 1.22  	 \\ 
 && 	 278.44  	 & 	 1.92  	 & 	 1.91 	 & 	 1.96  	 & 	      	 & 	      	 & 	      	 & 	      	 & 	      	 & 	      	 & 	      	 \\ 
 && 	 273.63  	 & 	 2.27  	 & 	 2.30 	 & 	 2.32  	 & 	      	 & 	      	 & 	      	 & 	      	 & 	      	 & 	      	 & 	      	 \\ 
 && 	 264.02  	 & 	 3.54  	 & 	 3.51 	 & 	 3.51  	 & 	      	 & 	      	 & 	      	 & 	      	 & 	      	 & 	      	 & 	      	 \\ 
 && 	 254.41  	 & 	 6.28  	 & 	 6.33 	 & 	 6.42  	 & 	      	 & 	      	 & 	      	 & 	      	 & 	      	 & 	      	 & 	      	 \\ 
\hline
\pagebreak
\multirow{15}{*}{3}	&	\multirow{15}{*}{91.71}	
 & 	 293.15  	 & 	 1.23  	 & 	 1.23 	 & 	 1.22  	 & 	 1.23  	 & 	 1.22  	 & 	 1.22  	 & 	 1.22  	 & 	 1.24  	 & 	 1.23  	 & 	 1.23  	 \\ 
 && 	 278.38  	 & 	 1.94  	 & 	 1.94 	 & 	 1.90  	 & 	      	 & 	      	 & 	      	 & 	      	 & 	      	 & 	      	 & 	      	 \\ 
 && 	 273.63  	 & 	 2.31  	 & 	 2.30 	 & 	 2.30  	 & 	      	 & 	      	 & 	      	 & 	      	 & 	      	 & 	      	 & 	      	 \\ 
 && 	 264.12  	 & 	 3.50  	 & 	 3.51 	 & 	 3.54  	 & 	      	 & 	      	 & 	      	 & 	      	 & 	      	 & 	      	 & 	      	 \\ 
 && 	 254.61  	 & 	 6.33  	 & 	 6.38 	 & 	 6.35  	 & 	      	 & 	      	 & 	      	 & 	      	 & 	      	 & 	      	 & 	      	 \\ 
 && 	 249.86  	 & 	 9.52  	 & 	 9.39 	 & 	 9.48  	 & 	      	 & 	      	 & 	      	 & 	      	 & 	      	 & 	      	 & 	      	 \\ 
 && 	 248.90  	 & 	 10.29  	 & 	 10.37 	 & 	 10.42  	 & 	      	 & 	      	 & 	      	 & 	      	 & 	      	 & 	      	 & 	      	 \\ 
 && 	 247.95  	 & 	 11.55  	 & 	 11.40 	 & 	      	 & 	      	 & 	      	 & 	      	 & 	      	 & 	      	 & 	      	 & 	      	 \\ 
 && 	 247.48  	 & 	 11.98  	 & 	 12.19 	 & 	 12.22  	 & 	      	 & 	      	 & 	      	 & 	      	 & 	      	 & 	      	 & 	      	 \\ 
 && 	 247.00  	 & 	 12.94  	 & 	 12.91 	 & 	 12.70  	 & 	      	 & 	      	 & 	      	 & 	      	 & 	      	 & 	      	 & 	      	 \\ 
 && 	 246.53  	 & 	 13.54  	 & 	 13.58 	 & 	      	 & 	      	 & 	      	 & 	      	 & 	      	 & 	      	 & 	      	 & 	      	 \\ 
 && 	 246.05  	 & 	 14.47  	 & 	 14.17 	 & 	 14.24  	 & 	      	 & 	      	 & 	      	 & 	      	 & 	      	 & 	      	 & 	      	 \\ 
 && 	 245.58  	 & 	 14.97  	 & 	 14.97 	 & 	 14.86  	 & 	      	 & 	      	 & 	      	 & 	      	 & 	      	 & 	      	 & 	      	 \\ 
 && 	 245.10  	 & 	 16.09  	 & 	 16.17 	 & 	 15.94  	 & 	      	 & 	      	 & 	      	 & 	      	 & 	      	 & 	      	 & 	      	 \\ 
 && 	 244.63  	 & 	 17.09  	 & 	 16.90 	 & 	      	 & 	      	 & 	      	 & 	      	 & 	      	 & 	      	 & 	      	 & 	      	 \\ 
\hline 
\multirow{11}{*}{4}	&	\multirow{11}{*}{60.45}	
 & 	 293.15  	 & 	 1.15  	 & 	 1.17 	 & 	 1.15  	 & 	 1.14  	 & 	 1.14  	 & 	 1.15  	 & 	 1.14  	 & 	 1.16  	 & 	 1.16  	 & 	 1.14  	 \\ 
 && 	 278.42  	 & 	 1.79  	 & 	 1.77 	 & 	 1.78  	 & 	      	 & 	      	 & 	      	 & 	      	 & 	      	 & 	      	 & 	      	 \\ 
 && 	 268.84  	 & 	 2.57  	 & 	 2.51 	 & 	      	 & 	      	 & 	      	 & 	      	 & 	      	 & 	      	 & 	      	 & 	      	 \\ 
 && 	 264.04  	 & 	 3.15  	 & 	 3.14 	 & 	 3.21  	 & 	      	 & 	      	 & 	      	 & 	      	 & 	      	 & 	      	 & 	      	 \\ 
 && 	 259.25  	 & 	 4.09  	 & 	 4.06 	 & 	 3.89  	 & 	      	 & 	      	 & 	      	 & 	      	 & 	      	 & 	      	 & 	      	 \\ 
 && 	 256.37  	 & 	 4.81  	 & 	 4.83 	 & 	 4.85  	 & 	      	 & 	      	 & 	      	 & 	      	 & 	      	 & 	      	 & 	      	 \\ 
 && 	 254.46  	 & 	 5.44  	 & 	 5.44 	 & 	 5.42  	 & 	      	 & 	      	 & 	      	 & 	      	 & 	      	 & 	      	 & 	      	 \\ 
 && 	 252.54  	 & 	 6.20  	 & 	 6.13 	 & 	 6.23  	 & 	      	 & 	      	 & 	      	 & 	      	 & 	      	 & 	      	 & 	      	 \\ 
 && 	 250.62  	 & 	 7.17  	 & 	 7.20 	 & 	 7.20  	 & 	      	 & 	      	 & 	      	 & 	      	 & 	      	 & 	      	 & 	      	 \\ 
 && 	 248.71  	 & 	 8.56  	 & 	 8.44 	 & 	 8.45  	 & 	      	 & 	      	 & 	      	 & 	      	 & 	      	 & 	      	 & 	      	 \\ 
 && 	 247.75  	 & 	 9.16  	 & 	 9.30 	 & 	 9.36  	 & 	      	 & 	      	 & 	      	 & 	      	 & 	      	 & 	      	 & 	      	 \\ 
\hline
\multirow{11}{*}{5}	&	\multirow{11}{*}{86.76}	
  & 	 293.15  	 & 	 1.22  	 & 	 1.20 	 & 	 1.23  	 & 	 1.21  	 & 	 1.21  	 & 	 1.21  	 & 	 1.22  	 & 	 1.20  	 & 	 1.22  	 & 	 1.21  	 \\ 
 && 	 283.19  	 & 	 1.60  	 & 	 1.64 	 & 	 1.60  	 & 	      	 & 	      	 & 	      	 & 	      	 & 	      	 & 	      	 & 	      	 \\ 
 && 	 278.41  	 & 	 1.88  	 & 	 1.89 	 & 	 1.89  	 & 	      	 & 	      	 & 	      	 & 	      	 & 	      	 & 	      	 & 	      	 \\ 
 && 	 273.63  	 & 	 2.28  	 & 	 2.28 	 & 	      	 & 	      	 & 	      	 & 	      	 & 	      	 & 	      	 & 	      	 & 	      	 \\ 
 && 	 268.85  	 & 	 2.73  	 & 	 2.72 	 & 	 2.74  	 & 	      	 & 	      	 & 	      	 & 	      	 & 	      	 & 	      	 & 	      	 \\ 
 && 	 264.07  	 & 	 3.44  	 & 	 3.47 	 & 	 3.44  	 & 	      	 & 	      	 & 	      	 & 	      	 & 	      	 & 	      	 & 	      	 \\ 
 && 	 259.29  	 & 	 4.48  	 & 	 4.52 	 & 	 4.52  	 & 	      	 & 	      	 & 	      	 & 	      	 & 	      	 & 	      	 & 	      	 \\ 
 && 	 256.42  	 & 	 5.39  	 & 	 5.39 	 & 	 5.36  	 & 	      	 & 	      	 & 	      	 & 	      	 & 	      	 & 	      	 & 	      	 \\ 
 && 	 254.51  	 & 	 6.10  	 & 	 6.16 	 & 	 6.18  	 & 	      	 & 	      	 & 	      	 & 	      	 & 	      	 & 	      	 & 	      	 \\ 
 && 	 252.60  	 & 	 7.07  	 & 	 7.07 	 & 	      	 & 	      	 & 	      	 & 	      	 & 	      	 & 	      	 & 	      	 & 	      	 \\ 
 && 	 250.68  	 & 	 8.45  	 & 	 8.39 	 & 	 8.47  	 & 	      	 & 	      	 & 	      	 & 	      	 & 	      	 & 	      	 & 	      	 \\ 
\hline
\pagebreak
\multirow{7}{*}{6}	&	\multirow{7}{*}{96.66}	
 & 	 293.15  	 & 	 1.25  	 & 	 1.25 	 & 	 1.24  	 & 	 1.23  	 & 	 1.24  	 & 	 1.23  	 & 	 1.23  	 & 	 1.24  	 & 	 1.24  	 & 	 1.24  	 \\ 
 && 	 278.64  	 & 	 1.94  	 & 	 1.93 	 & 	 1.92  	 & 	      	 & 	      	 & 	      	 & 	      	 & 	      	 & 	      	 & 	      	 \\ 
 && 	 269.18  	 & 	 2.85  	 & 	 2.92 	 & 	 2.92  	 & 	      	 & 	      	 & 	      	 & 	      	 & 	      	 & 	      	 & 	      	 \\ 
 && 	 264.45  	 & 	 3.56  	 & 	 3.56 	 & 	 3.53  	 & 	      	 & 	      	 & 	      	 & 	      	 & 	      	 & 	      	 & 	      	 \\ 
 && 	 259.72  	 & 	 4.75  	 & 	 4.70 	 & 	 4.63  	 & 	      	 & 	      	 & 	      	 & 	      	 & 	      	 & 	      	 & 	      	 \\ 
 && 	 256.88  	 & 	 5.70  	 & 	 5.63 	 & 	 5.63  	 & 	      	 & 	      	 & 	      	 & 	      	 & 	      	 & 	      	 & 	      	 \\ 
 && 	 254.99  	 & 	 6.54  	 & 	 6.45 	 & 	 6.37  	 & 	      	 & 	      	 & 	      	 & 	      	 & 	      	 & 	      	 & 	      	 \\ 
\hline
\multirow{15}{*}{7}	&	\multirow{15}{*}{96.66}
 & 	 293.15  	 & 	 1.24  	 & 	 1.25 	 & 	 1.24  	 & 	 1.24  	 & 	 1.23  	 & 	 1.24  	 & 	 1.23  	 & 	 1.24  	 & 	 1.24  	 & 	 1.23  	 \\ 
 && 	 278.64  	 & 	 1.93  	 & 	 1.94 	 & 	 1.96  	 & 	      	 & 	      	 & 	      	 & 	      	 & 	      	 & 	      	 & 	      	 \\ 
 && 	 269.18  	 & 	 2.87  	 & 	 2.86 	 & 	 2.88  	 & 	      	 & 	      	 & 	      	 & 	      	 & 	      	 & 	      	 & 	      	 \\ 
 && 	 264.45  	 & 	 3.67  	 & 	 3.67 	 & 	 3.67  	 & 	      	 & 	      	 & 	      	 & 	      	 & 	      	 & 	      	 & 	      	 \\ 
 && 	 259.72  	 & 	 4.73  	 & 	 4.72 	 & 	 4.83  	 & 	      	 & 	      	 & 	      	 & 	      	 & 	      	 & 	      	 & 	      	 \\ 
 && 	 256.88  	 & 	 5.85  	 & 	 5.80 	 & 	 5.78  	 & 	      	 & 	      	 & 	      	 & 	      	 & 	      	 & 	      	 & 	      	 \\ 
 && 	 254.99  	 & 	 6.62  	 & 	 6.60 	 & 	 6.54  	 & 	      	 & 	      	 & 	      	 & 	      	 & 	      	 & 	      	 & 	      	 \\ 
 && 	 253.10  	 & 	 7.47  	 & 	 7.67 	 & 	 7.58  	 & 	      	 & 	      	 & 	      	 & 	      	 & 	      	 & 	      	 & 	      	 \\ 
 && 	 251.21  	 & 	 8.89  	 & 	 8.84 	 & 	 8.91  	 & 	      	 & 	      	 & 	      	 & 	      	 & 	      	 & 	      	 & 	      	 \\ 
 && 	 249.32  	 & 	 10.51  	 & 	 10.68 	 & 	 10.70  	 & 	      	 & 	      	 & 	      	 & 	      	 & 	      	 & 	      	 & 	      	 \\ 
 && 	 248.37  	 & 	 11.69  	 & 	 11.97 	 & 	 11.85  	 & 	      	 & 	      	 & 	      	 & 	      	 & 	      	 & 	      	 & 	      	 \\ 
 && 	 247.43  	 & 	 13.13  	 & 	 13.14 	 & 	 12.94  	 & 	      	 & 	      	 & 	      	 & 	      	 & 	      	 & 	      	 & 	      	 \\ 
 && 	 246.48  	 & 	 14.80  	 & 	 14.62 	 & 	 14.77  	 & 	      	 & 	      	 & 	      	 & 	      	 & 	      	 & 	      	 & 	      	 \\ 
 && 	 245.54  	 & 	 16.47  	 & 	 16.75 	 & 	 16.52  	 & 	      	 & 	      	 & 	      	 & 	      	 & 	      	 & 	      	 & 	      	 \\ 
 && 	 244.59  	 & 	 18.50  	 & 	 18.70 	 & 	 18.59  	 & 	      	 & 	      	 & 	      	 & 	      	 & 	      	 & 	      	 & 	      	 \\ 
\hline
\multirow{11}{*}{8}	&	\multirow{11}{*}{52.77}	
  & 	 293.15  	 & 	 1.16  	 & 	 1.13 	 & 	 1.14  	 & 	 1.13  	 & 	 1.12  	 & 	 1.13  	 & 	 1.12  	 & 	 1.13  	 & 	 1.12  	 & 	 1.13  	 \\ 
 && 	 278.61  	 & 	 1.79  	 & 	 1.70 	 & 	 1.73  	 & 	      	 & 	      	 & 	      	 & 	      	 & 	      	 & 	      	 & 	      	 \\ 
 && 	 269.03  	 & 	 2.48  	 & 	 2.58 	 & 	 2.60  	 & 	      	 & 	      	 & 	      	 & 	      	 & 	      	 & 	      	 & 	      	 \\ 
 && 	 264.24  	 & 	 3.13  	 & 	 3.04 	 & 	      	 & 	      	 & 	      	 & 	      	 & 	      	 & 	      	 & 	      	 & 	      	 \\ 
 && 	 259.44  	 & 	 3.96  	 & 	 3.92 	 & 	      	 & 	      	 & 	      	 & 	      	 & 	      	 & 	      	 & 	      	 & 	      	 \\ 
 && 	 256.57  	 & 	 4.63  	 & 	 4.48 	 & 	      	 & 	      	 & 	      	 & 	      	 & 	      	 & 	      	 & 	      	 & 	      	 \\ 
 && 	 254.65  	 & 	 5.38  	 & 	 5.24 	 & 	 5.26  	 & 	      	 & 	      	 & 	      	 & 	      	 & 	      	 & 	      	 & 	      	 \\ 
 && 	 252.73  	 & 	 5.90  	 & 	 5.88 	 & 	      	 & 	      	 & 	      	 & 	      	 & 	      	 & 	      	 & 	      	 & 	      	 \\ 
 && 	 250.82  	 & 	 6.88  	 & 	     	 & 	      	 & 	      	 & 	      	 & 	      	 & 	      	 & 	      	 & 	      	 & 	      	 \\ 
 && 	 248.90  	 & 	 7.90  	 & 	 7.93 	 & 	      	 & 	      	 & 	      	 & 	      	 & 	      	 & 	      	 & 	      	 & 	      	 \\ 
 && 	 247.94  	 & 	 8.43  	 & 	     	 & 	      	 & 	      	 & 	      	 & 	      	 & 	      	 & 	      	 & 	      	 & 	      	 \\ 
\hline
\pagebreak
\multirow{9}{*}{9}	&	\multirow{9}{*}{49.96}	
  & 	 293.15  	 & 	 1.12  	 & 	 1.12 	 & 	 1.13  	 & 	 1.12  	 & 	 1.14  	 & 	 1.13  	 & 	 1.12  	 & 	 1.12  	 & 	 1.13  	 & 	 1.11  	 \\ 
 && 	 278.58  	 & 	 1.72  	 & 	 1.74 	 & 	      	 & 	      	 & 	      	 & 	      	 & 	      	 & 	      	 & 	      	 & 	      	 \\ 
 && 	 269.05  	 & 	 2.43  	 & 	 2.49 	 & 	 2.47  	 & 	      	 & 	      	 & 	      	 & 	      	 & 	      	 & 	      	 & 	      	 \\ 
 && 	 264.28  	 & 	 3.09  	 & 	 3.13 	 & 	 3.03  	 & 	      	 & 	      	 & 	      	 & 	      	 & 	      	 & 	      	 & 	      	 \\ 
 && 	 259.52  	 & 	 4.02  	 & 	 3.91 	 & 	 4.04  	 & 	      	 & 	      	 & 	      	 & 	      	 & 	      	 & 	      	 & 	      	 \\ 
 && 	 256.66  	 & 	 4.58  	 & 	     	 & 	      	 & 	      	 & 	      	 & 	      	 & 	      	 & 	      	 & 	      	 & 	      	 \\ 
 && 	 254.75  	 & 	 5.25  	 & 	 5.19 	 & 	 5.06  	 & 	      	 & 	      	 & 	      	 & 	      	 & 	      	 & 	      	 & 	      	 \\ 
 && 	 252.84  	 & 	 6.04  	 & 	 5.83 	 & 	 5.96  	 & 	      	 & 	      	 & 	      	 & 	      	 & 	      	 & 	      	 & 	      	 \\ 
 && 	 250.94  	 & 	 6.77  	 & 	     	 & 	      	 & 	      	 & 	      	 & 	      	 & 	      	 & 	      	 & 	      	 & 	      	 \\ 
\hline
\multirow{15}{*}{10}	&	\multirow{15}{*}{47.46}	
 & 	 293.15  	 & 	 1.14  	 & 	 1.15 	 & 	 1.14  	 & 	 1.11  	 & 	 1.13  	 & 	 1.12  	 & 	 1.08  	 & 	 1.09  	 & 	 1.11  	 & 	 1.11  	 \\ 
 && 	 278.68  	 & 	 1.69  	 & 	 1.69 	 & 	 1.68  	 & 	      	 & 	      	 & 	      	 & 	      	 & 	      	 & 	      	 & 	      	 \\ 
 && 	 269.15  	 & 	 2.41  	 & 	 2.38 	 & 	 2.40  	 & 	      	 & 	      	 & 	      	 & 	      	 & 	      	 & 	      	 & 	      	 \\ 
 && 	 264.38  	 & 	 2.95  	 & 	 2.96 	 & 	 2.94  	 & 	      	 & 	      	 & 	      	 & 	      	 & 	      	 & 	      	 & 	      	 \\ 
 && 	 259.61  	 & 	 3.81  	 & 	 3.78 	 & 	 3.79  	 & 	      	 & 	      	 & 	      	 & 	      	 & 	      	 & 	      	 & 	      	 \\ 
 && 	 256.75  	 & 	 4.47  	 & 	 4.53 	 & 	 4.49  	 & 	      	 & 	      	 & 	      	 & 	      	 & 	      	 & 	      	 & 	      	 \\ 
 && 	 254.84  	 & 	 5.00  	 & 	 4.92 	 & 	 5.02  	 & 	      	 & 	      	 & 	      	 & 	      	 & 	      	 & 	      	 & 	      	 \\ 
 && 	 252.94  	 & 	 5.59  	 & 	 5.64 	 & 	 5.52  	 & 	      	 & 	      	 & 	      	 & 	      	 & 	      	 & 	      	 & 	      	 \\ 
 && 	 251.03  	 & 	 6.50  	 & 	 6.47 	 & 	 6.66  	 & 	      	 & 	      	 & 	      	 & 	      	 & 	      	 & 	      	 & 	      	 \\ 
 && 	 249.12  	 & 	 7.51  	 & 	 7.49 	 & 	 7.53  	 & 	      	 & 	      	 & 	      	 & 	      	 & 	      	 & 	      	 & 	      	 \\ 
 && 	 248.17  	 & 	 8.18  	 & 	 8.12 	 & 	 8.25  	 & 	      	 & 	      	 & 	      	 & 	      	 & 	      	 & 	      	 & 	      	 \\ 
 && 	 247.22  	 & 	 8.90  	 & 	 8.87 	 & 	 8.74  	 & 	      	 & 	      	 & 	      	 & 	      	 & 	      	 & 	      	 & 	      	 \\ 
 && 	 246.26  	 & 	 9.61  	 & 	 9.61 	 & 	 9.58  	 & 	      	 & 	      	 & 	      	 & 	      	 & 	      	 & 	      	 & 	      	 \\ 
 && 	 245.31  	 & 	 10.63  	 & 	 10.60 	 & 	 10.48  	 & 	      	 & 	      	 & 	      	 & 	      	 & 	      	 & 	      	 & 	      	 \\ 
 && 	 244.36  	 & 	 11.75  	 & 	     	 & 	      	 & 	      	 & 	      	 & 	      	 & 	      	 & 	      	 & 	      	 & 	      	 \\ 
\hline
\end{longtable*}
\end{widetext}

%